\newif\iftwocol
\twocoltrue % comment if single column

\iftwocol
\documentclass[journal,twocolumn,final]{IEEEtran}
\else
\documentclass[journal,onecolumn,draftcls,12pt]{IEEEtran}
\fi

\usepackage{header}
\usepackage[multiple]{footmisc}
\usepackage[utf8]{inputenc}

\makeatletter
\def\@IEEEinterspaceratioM{0.265}
\def\@IEEEinterspaceMINratioM{0.1651}
\def\@IEEEinterspaceMAXratioM{0.38}

\def\@IEEEinterspaceratioB{0.31}
\def\@IEEEinterspaceMINratioB{0.19}
\def\@IEEEinterspaceMAXratioB{0.38}
\@IEEEtunefonts
\makeatother
\begin{document}

\title{Finite-SNR Bounds on the Sum-Rate Capacity of  Rayleigh Block-Fading Multiple-Access Channels with no a Priori  CSI}

\author{Rahul~Devassy,~\IEEEmembership{Student Member,~IEEE,}
		Giuseppe~Durisi,~\IEEEmembership{Senior Member,~IEEE,}
        Johan~Östman, %~\IEEEmembership{Member,~IEEE,}
        Wei~Yang,~\IEEEmembership{Student Member,~IEEE,}
        Tome~Eftimov, %~\IEEEmembership{Student Member,~IEEE,}
        Zoran~Utkovski,~\IEEEmembership{Member,~IEEE} 

 \thanks{This work was supported in part by the Swedish Research Council under grant 2012-4571 and in part by the German Research
Society (Deutsche Forschungsgemeinschaft), via the project Li 659/13-1. 
 The simulations were performed on resources at Chalmers Centre of Computational Science and Engineering (C3SE) provided by the Swedish National Infrastructure for Computing (SNIC).}
 \thanks{R. Devassy,  G. Durisi and W. Yang are with the Department of Signals and Systems, Chalmers University of Technology, Gothenburg, Sweden (e-mail: \{devassy,durisi,ywei\}@chalmers.se).}
 \thanks{J. \"Ostman is with the Department of Electronic and Computer Engineering, Hong Kong University of Science and Technology, Hong Kong, China (email: eejohan@ust.hk).}
 \thanks{T. Eftimov is with the Laboratory for Complex
Systems and Networks, Macedonian Academy of Sciences and Arts, R. Macedonia. (email: {tomeeftimov@gmail.com}).}
 \thanks{Z. Utkovski is with the Faculty of Computer Science, University Goce Delčev Štip, and with the Laboratory for Complex Systems and Networks, Macedonian
Academy of Sciences and Arts, R. Macedonia.  (e-mail: zoran.utkovski@ugd.edu.mk).}
 
}

\maketitle

%%%%%%%%%%%%%%%%
\begin{abstract}
We provide nonasymptotic upper and lower bounds on the sum-rate capacity of Rayleigh block-fading multiple-access channels for the set up where \apriori channel state information is not available. The upper bound relies on a dual formula for channel capacity and on the assumption that the users can cooperate perfectly.
The lower bound is derived assuming a noncooperative scenario where each user employs unitary space-time modulation (independently from the other users). Numerical results show that the gap between the upper and the lower bound is small already at moderate SNR values. This suggests that the sum-rate capacity gains obtainable through user cooperation are minimal \marktext{for the scenarios considered in the paper.}
\end{abstract}

\section{Introduction} 
\label{sec_introduction}
The multiple-access channel (MAC) models a scenario where two or more noncooperating users communicate with a single receiver. This scenario is relevant for the uplink of wireless cellular networks, where the users may be mobile terminals and the receiver may be a cellular base station.  In this paper, we consider the setup where neither the users nor the receiver have \apriori information on the realization of the fading process. 
Such a situation arises in high mobility scenarios where it is not desirable for the receiver 
to feed back channel state information (CSI) to the users because it may be outdated~\cite{5374062}. It may also arise in the initialization phase of a communication link, e.g., when a mobile terminal joins a cellular network. 
Capacity analyses under assumption of no \apriori CSI have the advantage of capturing the cost of estimating the fading channel, and, hence, yield more realistic throughput estimates than analyses based on the assumptions of perfect CSI~\cite{lapidoth2005asymptotic,moser2009fading,yang2012diversity}. %TBD : reference [also refer to papers that analyze fading MAC with CSIR?].
Indeed, under the assumption of no feedback from the receiver to the user terminals, these capacity analyses provide a fundamental limit on the performance of every communication scheme, irrespectively of whether it relies on explicit channel estimation or not. 

In this paper, we consider the case where  the users as well as the receiver are equipped with one or more antennas. 
We shall focus on the so-called Rayleigh block-fading model~\cite{825818,marzetta1999capacity}. The two key features of this model are that i) the fading coefficients associated to the channels between each transmit and receive antenna pair are independent and identically distributed (\iid) circularly symmetric complex Gaussian random variables; ii) each fading coefficient remains constant over $\cohtime$ channel uses before changing to a new independent realization. The parameter~$\cohtime,$ which is the ratio between the channel coherence time and the symbol duration, will be referred to in this paper as \emph{coherence interval}.

The capacity of Rayleigh block-fading channels under the assumption of no \apriori CSI has been studied extensively in the  point-to-point case. Specifically, considering a system with $\numtxant$ transmit and $\numrxant$ receive antennas, Marzetta and Hochwald proved that the capacity-achieving input matrix $\cinp\in\complexset^{\numtxant\times\cohtime}$ can be expressed as~\cite[Thm. 2]{marzetta1999capacity}
\begin{IEEEeqnarray}{rCl}
\cinp &=& \bD \bQ \label{marzetta_xdphi}
\end{IEEEeqnarray}
where $\bD$ is a real $\numtxant\times\numtxant$ diagonal matrix whose diagonal entries have a joint probability density function (p.d.f.) that is invariant to permutation of its arguments, and $\bQ\in\complexset^{\numtxant\times\cohtime},$ independent of $\bD,$ is an isotropically distributed matrix with orthonormal rows (truncated unitary matrix).

In spite of the partial characterization of the capacity-achieving input distribution provided in~\cite[Thm. 2]{marzetta1999capacity}, no closed form expression for capacity is available to date. However, the high-SNR capacity behavior is well understood. Indeed, extending a result obtained for the single-input single-output case in~\cite{825818}, Zheng and Tse~\cite{zheng2002communication} proved that in the high-SNR regime, the capacity~$\capacity$ of a $\numtxant\times\numrxant$ multiple-input multiple-output (MIMO) Rayleigh block-fading channel behaves as \begin{IEEEeqnarray}{rCl}
\capacity(\snr) &=& n^{*}\left(1-\frac{n^{*}}{\cohtime}\right)\Blog{\snr}+\bigo{1}.\label{zheng_MIMO_cap}
\end{IEEEeqnarray} 
Here, $\snr$ stands for the SNR, $n^{*} = \min(\numtxant,\numrxant,\floor{\cohtime/2}),$ and $\bigo{1}$ indicates a function whose magnitude is upper-bounded by a finite constant for sufficiently large SNR values. The asymptotic expression~\eqref{zheng_MIMO_cap} can be tightened to~\cite{zheng2002communication,6415385}\begin{IEEEeqnarray}{rCl}
\capacity(\snr) &=&
 n^{*}\left(1-\frac{n^{*}}{\cohtime}\right)\Blog{\snr}+c+\smallo{1} \label{zheng_MIMO_sp_cap}
\end{IEEEeqnarray} where $c$ is the constant given in~\cite[Eq. (9)]{6415385} and $\smallo{1}$ denotes a function that vanishes as~$\snr\rightarrow\infty.$ For the case when $\cohtime\geq\numtxant+\numrxant,$ one can achieve~\eqref{zheng_MIMO_sp_cap} by choosing $\bD$ in~\eqref{marzetta_xdphi} to be a scaled identity matrix~\cite{zheng2002communication}. The resulting probability distribution on $\cinp$ is commonly referred to as unitary space-time modulation (USTM)~\cite{825818}. When $\cohtime<\numtxant+\numrxant,$ the matrix $\bD$ must be chosen so that its diagonal entries are distributed as the square root of the eigenvalues of a Beta-distributed random matrix of appropriate size~\cite{6415385}. The resulting probability distribution on $\cinp$ is referred to in~\cite{6415385} as Beta-variate space-time modulation (BSTM).

Nonasymptotic (i.e., finite-SNR) lower bounds on the capacity of point-to-point MIMO Rayleigh block-fading channels have been obtained for specific probability distributions on $\cinp.$ Specifically, an \iid Gaussian input distribution is considered in~\cite{rusek2012mutual}, USTM in~\cite{hassibi2002multiple}, and BSTM in~\cite{alfano2014closed}, where the analysis is also extended to Rician block-fading and to land mobile satellite channels. A key tool in the derivation of these nonasymptotic lower bounds is the Itzykson-Zuber integral~\cite[Eq. (3.4)]{itzykson1980planar}, which allows one to obtain a closed-form expression for the conditional probability distribution of the channel output given the diagonal input matrix $\bD$ in~\eqref{marzetta_xdphi}. The method employed so far to assess the tightness of the bounds obtained in~\cite{rusek2012mutual,hassibi2002multiple,alfano2014closed} is to compare these lower bounds with the asymptotic expansion~\eqref{zheng_MIMO_sp_cap} (with the $\smallo{1}$ term omitted). Unfortunately, this method is not conclusive because the error incurred by omitting the $\smallo{1}$ term in~\eqref{zheng_MIMO_sp_cap} is not quantified. A simple capacity upper bound can be obtained by assuming that a genie provides the receiver with perfect CSI. However, this bound is tight only when $\cohtime$ is large and the channel estimation overhead negligible.

Leaving the point-to-point case and moving to the MAC, we note that the independence constraint on the signals transmitted by the various users implies that the partial characterization of the capacity-achieving input distribution obtained for the point-to-point case in~\cite[Thm. 2]{marzetta1999capacity}, as well as the asymptotic capacity expansions in~\cite{zheng2002communication,6415385} and the nonasymptotic capacity lower bounds in~\cite{hassibi2002multiple,alfano2014closed} do not carry over to the MAC sum-rate capacity. The only exception is the \iid Gaussian lower bound obtained in~\cite{rusek2012mutual}, which also applies to the MAC because the transmission of \iid Gaussian signals does not require coordination among the users.

Coarse upper and lower bounds on the MAC sum-rate capacity for the case when $\numuser$ single-antenna users communicate with a receiver equipped with $\numrxant$ antennas are provided in~\cite{992788}. By examining these bounds in two different asymptotic regimes (high SNR and large $\cohtime,$ for a fixed $\numuser/\cohtime$ ratio) the authors conjecture that the sum-rate capacity is maximized when $\numuser=\cohtime.$ In the same paper, the authors pose the question of whether the constraint that the users transmit independent signals yields a sum-rate capacity \emph{prelog}\footnote{The \emph{prelog} (a.k.a. multiplexing gain) is the asymptotic ratio between the (sum-rate) capacity and $\Blog{\snr}$ in the limit~$\snr\rightarrow\infty.$} that is strictly lower than the one achievable when the users can cooperate perfectly, and the MAC reduces to a point-to-point MIMO channel, for which~\eqref{zheng_MIMO_cap} holds. It follows from~\cite[Sec. V]{zheng2002communication} that this is not the case, provided that the users are able to transmit orthogonal pilot signals, used at the receiver to estimate the channel.

Lin and Moser~\cite{5961836} characterized the high-SNR behavior of the sum-rate capacity of an \iid Rician-fading MAC (block-fading channel with coherence interval $\cohtime = 1$). They showed that the sum-rate capacity grows double-logarithmically in SNR, and that the sum-rate capacity maximizing strategy at high SNR is to switch off all users but one. 

\subsubsection*{Contributions}
We present nonasymptotic (i.e., finite-SNR) upper and lower bounds on the sum-rate capacity of Rayleigh block-fading MACs. 
\marktext{Similarly to the nonasymptotic capacity lower bounds previously reported in~\cite{rusek2012mutual,hassibi2002multiple,alfano2014closed}, our bounds are not in closed form, but they can be evaluated numerically.}
Our upper bound is obtained by assuming that the users can perfectly cooperate, which turns the MAC into an equivalent MIMO point-to-point channel. 
In addition, we use the duality upper bound on mutual information reported in~\cite[Eq. (186)]{lapidoth2003capacity}. 
As \emph{auxiliary} output distribution in the duality step, we choose the one induced by USTM inputs in the absence of additive noise. This method was used in~\cite{6415385} to establish the asymptotic expansion~\eqref{zheng_MIMO_sp_cap} for the case $\numtxant\leq \min(\numrxant,\floor{\cohtime/2}).$ Here, we provide a finite-SNR  analysis, which generalizes to MIMO the one reported in~\cite{yang2012diversity} for the single-input single-output (SISO) case.

% FIGURES for explaining contributions visually 
\begin{figure}
\centering
\includegraphics{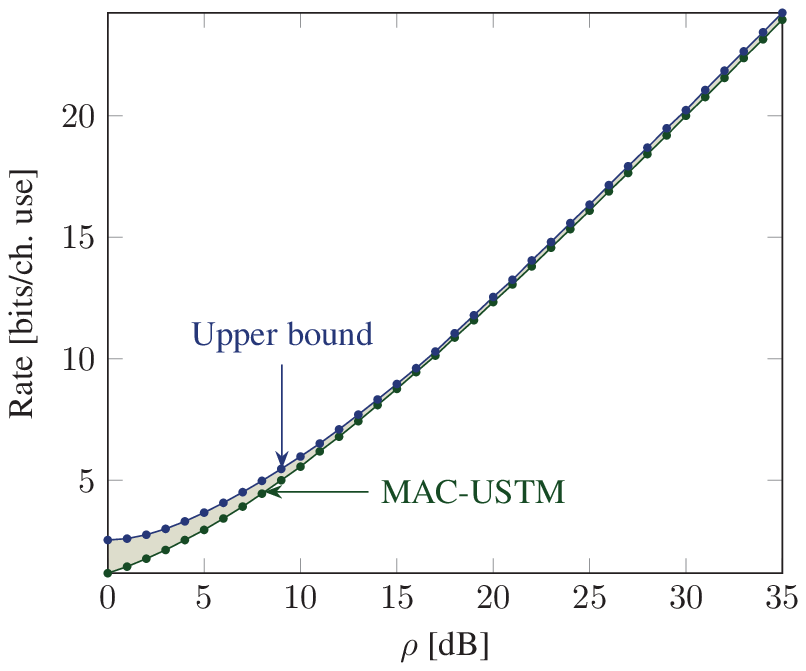}
\caption{Upper and lower bounds on the MAC sum-rate capacity: $4$ single-antenna users; receiver with $4$ antennas; coherence interval of $10$ channel uses.}\label{fig_intro_cap}
\end{figure}
% macustam vs gaussian
\begin{figure}
\centering
\includegraphics{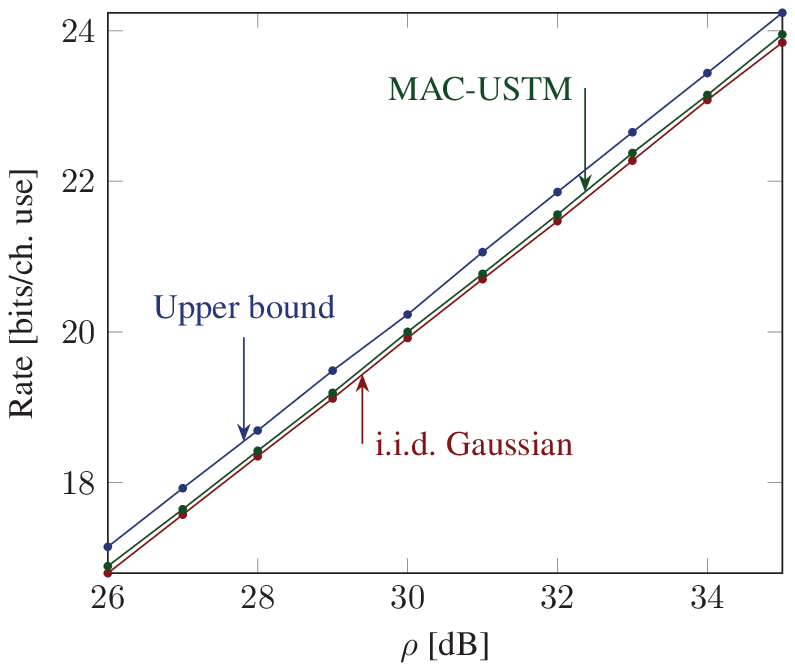}
\caption{\marktext{Upper bound, \sphuniform lower bound, and \iid Gaussian lower bound on the MAC sum-rate capacity}: $4$ single-antenna users; receiver with $4$ antennas; coherence interval of $10$ channel uses.}
\label{fig_intro_cmp}
\end{figure}

The nonasymptotic lower bound is obtained by allowing each user to transmit a USTM signal over the available antennas. We call the resulting input distribution \sphuniform. Note that \sphuniform does not yield a global USTM input distribution, because independence among users imply that orthogonality among them cannot be enforced. Numerical evidence (see Fig.~\ref{fig_intro_cap} and Fig.~\ref{fig_intro_cmp}) suggests that \marktext{for the scenarios considered in the paper}: \begin{itemize}
\item Our upper and lower bounds characterize accurately the sum-rate capacity (see Fig.~\ref{fig_intro_cap}).
\item The gain in sum-rate capacity obtainable by allowing user co-operation is minimal. This follows because the upper bound is obtained under the assumption of perfect cooperation between users.
A similar observation has been recently reported in~\cite{mendlovic14-06a} for the nonfading asynchronous Gaussian MAC.
\item Orthogonal pilot transmission is not required to obtain rates close to the full-cooperation case (cf.~\cite{992788},~\cite[Sec. V]{zheng2002communication}).
\item The \sphuniform lower bound is tighter than the \iid Gaussian lower bound, although the gain is marginal when the users are equipped with a single antenna (see Fig.~\ref{fig_intro_cmp}). This suggests that \iid Gaussian inputs are almost sum-rate capacity optimal already at moderate SNR values, which confirms an observation reported in~\cite{rusek2012mutual} for the point-to-point MIMO case.
\end{itemize}

\subsubsection*{Notation} Uppercase letters denote matrices,\footnote{We do not distinguish between vectors and matrices. We treat $n$-dimensional vectors as $1\times n$ or $n\times 1$ matrices.} lowercase letters designate scalars, and boldface letters denote random quantities. 
The superscript$~\dag$ stands for Hermitian transposition. 
For a random variable~$\bmx$ with p.d.f.~$f_{\bmx}(x),$ we write $\bmx\sim f_{\bmx}(x).$ When two random variables $\bmx$ and $\bmy$ have the same p.d.f., we write $\bmx\sim \bmy.$ For a matrix $X,$ we let~$X_{ij}$ denote its entry on the $i$th row and the $j$th column. With $\Bdet{f(i,j)}$ we indicate the determinant of a matrix whose entry on the $i$th row and the $j$th column is given by $f(i,j),$ for some arbitrary function $f(\cdot,\cdot).$ It will turn out convenient to define the following function
\begin{IEEEeqnarray}{c}
\gammas{n} = \left\{ \,
\begin{IEEEeqnarraybox}[][c]{l?l}
\IEEEstrut
\prod_{i=1}^n\Gamma(i), & n\in \naturals \\
1, & n = 0
\IEEEstrut
\end{IEEEeqnarraybox}
\right.
\label{gammas}
\end{IEEEeqnarray}
  where $\Gamma(\cdot)$ is the Gamma function~\cite[Sec. 6.1]{abramowitz2012handbook}. With $\beta(\cdot,\cdot)$ we denote the Beta function~\cite[Sec. 6.2]{abramowitz2012handbook} and $\psi(\cdot)$ denotes the Digamma function~\cite[Sec. 6.3]{abramowitz2012handbook}. The identity matrix of dimension $n\times n$ is denoted by $I_n.$ We let $\DMat{d_1,d_2,\dots,d_n}$ be the diagonal matrix with entries~$d_1,d_2,\dots,d_n$ on its diagonal. For a Hermitian matrix $A\in \complexset^{n\times n}$ with ordered eigenvalues~$a_1> a_2 > \dots > a_n$ we denote the determinant of the Vandermonde matrix constructed from~$a_1,a_2,\dots,a_n$ as
\begin{IEEEeqnarray}{c}
\vand(A)=\prod_{1\leq i<j\leq n}(a_i-a_j). \label{vanderdet}
\end{IEEEeqnarray}
We shall often use the following two functions
\begin{IEEEeqnarray}{c}
\kappas{A}{k} = \vand(A)\Bdet{A}^{k}
\label{kappadef}
\end{IEEEeqnarray} where $A\in \complexset^{n\times n}$ is Hermitian and $k\in\naturals,$ and
\begin{IEEEeqnarray}{rCl}
\Bupperexp{x}{n} &=&  e^x-\sum_{k=0}^{n-1}\frac{x^k}{k!} \label{gamma_x_n_def}
\label{incgammadef}
\end{IEEEeqnarray} where $x\in\reals$ and $n\in\naturals.$
The set of unitary matrices in $\complexset^{n\times n}$ is denoted by $\unitgrp(n)$ (\emph{unitary group}) and the set of matrices $U\in \complexset^{n\times m},m\geq n$ with $UU^\dag = I_n$ is denoted by $\isotropicgrp(n,m)$ (\emph{\marktext{Stiefel} manifold}). With $\Bexoprand{\bmx}{f(\bmx)},$ we denote the expectation of the function $f(\bmx)$ over the random variable $\bmx$. 
We let $\difent(\bmx)$ denote the differential entropy of a continuous random variable~$\bmx$; furthermore, $\mutualinfo{\bmy}{\bmz}$ stands for the mutual information between the random variables $\bmy$ and $\bmz.$ The set of all diagonal matrices in $\reals^{n\times m}$ with ordered \marktext{and distinct} positive entries on their main diagonal is denoted by~$\diagmatset{n}{m}.$ With $\zeromat{n}{m}$ we indicate the $n\times m$ zero matrix. We use $\cnormdist{0}{\sigma^2}$ to denote a circularly symmetric complex Gaussian random variable with zero mean and variance $\sigma^2,$ and $\betadist{a}{b}$ to denote a Beta-distributed random variable with parameters $a$ and $b.$

\section{System Model}
We consider a Rayleigh block-fading MAC where $\numuser$ users communicate with a receiver having~$\numrxant$ antennas, and the channel coherence interval is $\cohtime$ (same for all users, which corresponds to a scenario where users with similar mobility requirements are scheduled together). We assume that each user is equipped with one or more antennas and denote by $\ithuserant$ the number of antennas at user $i,\ i=1,\dots,\numuser.$ The received signal $\cout\in\complexset^{\numrxant\times\cohtime}$ over an arbitrary coherence interval can be compactly written in matrix notation as follows:
\begin{IEEEeqnarray}{c}
\cout=\sum_{i=1}^{\numuser}\channel_i\cinp_i+\noise. \label{ysixiw}
\end{IEEEeqnarray}
Here, $\cinp_i\in\complexset^{\ithuserant\times\cohtime}$ denotes the signal transmitted by user $i$ over the coherence interval, and the matrix~$\channel_i \in \complexset^{\numrxant\times \ithuserant}$ contains the fading coefficients associated to the channels between each transmit antenna of user $i$ and the receive antennas, within the coherence interval. We assume that $\channel_i$ has \iid $\cnormdist{0}{1}$ entries and that the channel matrices $\listofvars{\channel_i}{i=1}{\numuser}$ are independent. Finally, the matrix~$\noise\in\complexset^{\numrxant\times\cohtime},$ whose entries are \iid~$\cnormdist{0}{1}$-distributed, denotes the additive noise. Let \begin{IEEEeqnarray}{c}
\numtxant = \sum_{i=1}^{\numuser} \ithuserant
\end{IEEEeqnarray} be the total number of transmit antennas. We can rewrite~\eqref{ysixiw} as \begin{IEEEeqnarray}{c}
\cout=\channel\cinp+\noise
\label{ysxw}
\end{IEEEeqnarray} where \begin{IEEEeqnarray}{C}
\channel = \begin{bmatrix}
\channel_1&\channel_2&\cdots&\channel_{\numuser}
\end{bmatrix} \in\complexset^{\numrxant\times\numtxant}\label{sdef}
\end{IEEEeqnarray} and  \begin{IEEEeqnarray}{C}
\cinp = \begin{bmatrix}
\cinp_1 \\
\cinp_2 \\
\vdots \\
\cinp_{\numuser}
\end{bmatrix}\in\complexset^{\numtxant\times\cohtime}. \label{xdef}
\end{IEEEeqnarray} We assume that $\noise$ and $\channel$ are independent, and that their probability law does not depend on $\cinp.$ Throughout the paper, we also assume that $\cohtime\geq\max(\numtxant,\numrxant)$ and focus on the scenario where neither the transmitter nor the receiver have prior knowledge of matrix $\channel$ (no \apriori CSI). It will turn out convenient to define the following two constants
\begin{IEEEeqnarray}{rCl}
\numminant &=& \min(\numtxant,\numrxant)\label{ldef}\\
\nummaxant &=& \max(\numtxant,\numrxant).\label{pdef}
\end{IEEEeqnarray}

The sum-rate capacity of the MAC in~\eqref{ysxw} is given by
\begin{IEEEeqnarray}{c}
\capacity(\snr) = \frac{1}{\cohtime} \sup_{} \mutualinfo{\cinp}{\cout} \label{sumratecap}
\end{IEEEeqnarray}
where the supremum is over all probability distributions on $\cinp$ for which 
\begin{enumerate}
\item $\listofvars{\cinp_i}{i=1}{\numuser}$ are independent;
\item the per-user power constraint \begin{IEEEeqnarray}{c}
\Bexop{\Btrace{\cinp_i\cinp_i^{\dag}}} \leq \frac{\cohtime \ithuserant \snr}{\numtxant}, \  i=1,2,\dots,\numuser \label{userpwrconst}
\end{IEEEeqnarray} is satisfied.
\end{enumerate}
Here, $\snr$ can be thought of as the total energy per channel use available over all users. The particular form of average power constraint in~\eqref{userpwrconst} allows all users to transmit at the same average power per antenna. As reviewed in Section \ref{sec_introduction}, $\capacity(\snr)$ is not known in closed form.

In the next section, we provide nonasymptotic upper and lower bounds on the sum-rate capacity $\capacity(\snr)$  in~\eqref{sumratecap}. These bounds will be numerically evaluated in Section~\ref{sec_sim_results}, and conclusions be drawn in Section~\ref{sec_conclusion}.

\section{Bounds on Capacity}
\label{sec_bounds}
Theorem \ref{theorem_ub} below provides a nonasymptotic upper bound on  $\capacity(\snr).$ The upper bound is obtained by dropping the requirement that the $\listofvars{\cinp_i}{i=1}{\numuser}$ are independent (which enlarges the set of distributions over which the maximization in~\eqref{sumratecap} is performed), and by bounding the mutual information in~\eqref{sumratecap} using the duality bound~\cite[Eq. (186)]{lapidoth2003capacity}. The duality approach requires the specification of an auxiliary p.d.f. on the channel output $\cout,$ which, following~\cite{6415385}, we choose so that i)~$\cout$ is isotropic; ii) the largest $\numminant$ singular values of $\cout$ are distributed as the singular values of the noiseless channel output $\channel\cinp,$ with $\cinp$ following a USTM distribution; iii) the remaining $\numrxant-\numminant$ singular values are distributed as the singular values of an additive noise matrix of appropriate dimension.

\begin{thm} \label{theorem_ub}
The sum-rate capacity $\capacity(\snr)$ in~\eqref{sumratecap} of the Rayleigh block-fading MAC~\eqref{ysixiw} is upper-bounded by
\begin{IEEEeqnarray}{rCl}
\capacity(\snr)  &\leq &  \marktext{u(\snr)}+ \frac{1}{\cohtime}\inf_{\lambda\geq 0}\sup_{D\in \diagmatset{\numtxant}{\numtxant}} g(D,\lambda). \label{ub_expr}
\end{IEEEeqnarray} 
Here, 
\marktext{
\iftwocol
\begin{IEEEeqnarray}{rCl}
\marktext{u(\snr)} & = &-\>\numrxant+\frac{\numrxant\numtxant}{\cohtime}\Blog{\frac{\cohtime\snr}{\numtxant}}  \nonumber\\
&&+\>\frac{1}{\cohtime}\Blog{\frac{\gammas{\numrxant-\numminant}\gammas{\cohtime-\numminant}\gammas{\numtxant}}{\gammas{\cohtime} \gammas{\nummaxant-\numminant}}}+ \frac{1}{\cohtime}\Blog{\orderconstant}\nonumber\\
 && +\>  \frac{\numtxant\numrxant}{\cohtime\snr}+ \frac{(\numrxant-\numminant)(\cohtime-\numminant)}{\cohtime} \nonumber\\
 &&-\> \frac{(\cohtime-\numtxant)(\numrxant-\numminant)}{\cohtime}\Blog{\expmaxaigenconstant}\IEEEeqnarraynumspace\label{ub_udef}
\end{IEEEeqnarray}
\else
\begin{IEEEeqnarray}{rCl}
\marktext{u(\snr)} & = &-\>\numrxant+\frac{\numrxant\numtxant}{\cohtime}\Blog{\frac{\cohtime\snr}{\numtxant}}  + \frac{1}{\cohtime}\Blog{\frac{\gammas{\numrxant-\numminant}\gammas{\cohtime-\numminant}\gammas{\numtxant}}{\gammas{\cohtime} \gammas{\nummaxant-\numminant}}}\nonumber\\
 && +\> \frac{1}{\cohtime}\Blog{\orderconstant} + \frac{\numtxant\numrxant}{\cohtime\snr}+ \frac{(\numrxant-\numminant)(\cohtime-\numminant)}{\cohtime}-\frac{(\cohtime-\numtxant)(\numrxant-\numminant)}{\cohtime}\Blog{\expmaxaigenconstant}\label{ub_udef}\end{IEEEeqnarray} 
\fi
}and 
\iftwocol
\begin{IEEEeqnarray}{rCl} 
g(D,\lambda )&=& \frac{\numtxant\numrxant\Btrace{D^{2}}}{\cohtime\snr} \nonumber \\
&&+\> (\cohtime-\numtxant)\Bexoprand{\bG}{\Blogdet{\bG(I_{\numtxant}+D^{2})\bG^{\dag}+\expmaxaigenconstant I_{\numrxant}}}\nonumber\\
&&-\> \numrxant \Blogdet{I_{\numtxant}+D^{2}} + \lambda(\cohtime\snr-\Btrace{D^{2}})\label{ub_gdef}
\end{IEEEeqnarray} 
\else
\begin{IEEEeqnarray}{rCl} 
g(D,\lambda )&=& \frac{\numtxant\numrxant\Btrace{D^{2}}}{\cohtime\snr}+ (\cohtime-\numtxant)\Bexoprand{\bG}{\Blogdet{\bG(I_{\numtxant}+D^{2})\bG^{\dag}+\expmaxaigenconstant I_{\numrxant}}}\nonumber\\
&&-\> \numrxant \Blogdet{I_{\numtxant}+D^{2}} + \lambda(\cohtime\snr-\Btrace{D^{2}})\label{ub_gdef}
\end{IEEEeqnarray} 
\fi
where $\gammas{\cdot}$ is given in~\eqref{gammas}, $\bG$ is an $\numrxant\times \numtxant$ complex random matrix with \iid $\cnormdist{0}{1}$ entries and $\expmaxaigenconstant$ is the expected value of the square of the largest singular value of an $\numrxant\times(\cohtime-\numtxant)$ complex random matrix with \iid $\cnormdist{0}{1}$ entries.  Finally, with $\orderconstant$ we denote the probability that the lowest nonzero singular value of an $\numrxant\times\numtxant$ complex random matrix with \iid $\cnormdist{0}{\cohtime\snr/\numtxant}$ entries is greater than the largest singular value of an independent $(\numrxant-\numminant)\times(\cohtime-\numminant)$ complex random matrix with  \iid $\cnormdist{0}{1}$ entries.
\end{thm}
\begin{IEEEproof} See Appendix \ref{appendix_ubproof}.\end{IEEEproof}

When $\numtxant=\numrxant,$ the sum-rate upper bound~\eqref{ub_expr} can be tightened. This result is given in the following corollary.
\begin{cor} \label{theorem_ub_sym}
The sum-rate capacity $\capacity(\snr)$ in~\eqref{sumratecap} of the Rayleigh block-fading MAC~\eqref{ysixiw} for the case~$\numtxant = \numrxant = \numsymant$ is upper-bounded by
\begin{IEEEeqnarray}{rCl}
\capacity(\snr)  &\leq & \marktext{u^{*}(\snr)}+ \frac{1}{\cohtime}\inf_{\lambda\geq 0}\sup_{D\in \diagmatset{\numsymant}{\numsymant}} g^{*}(D,\lambda). \label{expr_ub_sym}
\end{IEEEeqnarray} 
Here, 
\marktext{\iftwocol
\begin{IEEEeqnarray}{rCl}
\marktext{u^{*}(\snr)}&=&-\>\numsymant+\frac{\numsymant^{2}}{\cohtime}\Blog{\frac{\cohtime\snr}{\numsymant}} +\frac{\numsymant^{2}}{\cohtime\snr}\nonumber \\
&&+\>\frac{1}{\cohtime}\Blog{\frac{\gammas{\numsymant}\gammas{\cohtime-\numsymant}}{\gammas{\cohtime}}} \label{ub_udef_sym}
\end{IEEEeqnarray}
\else
\begin{IEEEeqnarray}{rCl}
\marktext{u^{*}(\snr)}&=&-\>\numsymant+\frac{\numsymant^{2}}{\cohtime}\Blog{\frac{\cohtime\snr}{\numsymant}} +\frac{\numsymant^{2}}{\cohtime\snr}+\frac{1}{\cohtime}\Blog{\frac{\gammas{\numsymant}\gammas{\cohtime-\numsymant}}{\gammas{\cohtime}}} \label{ub_udef_sym}
\end{IEEEeqnarray}
\fi }
and
\iftwocol
\begin{IEEEeqnarray}{rCl}
g^{*}(D,\lambda)&=&\frac{\numsymant^2\Btrace{D^{2}}}{\cohtime\snr}-\numsymant\Blogdet{I_{\numsymant}+D^{2}}\nonumber\\
&&+\> \lambda(\cohtime\snr-\Btrace{D^{2}})\nonumber\\
&&+\> \frac{(\cohtime-\numsymant) \sum_{k=1}^{\numsymant}\Bdet{R_{k}(I_{\numsymant}+D^{2})} }{\kappas{D^{2}}{\cohtime-\numsymant}\Bdet{I_{\numsymant}+D^{2}}^{-\cohtime+\numsymant+1}} \IEEEeqnarraynumspace
 \label{ub_gdef_sym}
\end{IEEEeqnarray} 
\else
\begin{IEEEeqnarray}{rCl}
g^{*}(D,\lambda)&=&\frac{\numsymant^2\Btrace{D^{2}}}{\cohtime\snr}-\numsymant\Blogdet{I_{\numsymant}+D^{2}} + \lambda(\cohtime\snr-\Btrace{D^{2}})\nonumber\\
&&+\> \frac{(\cohtime-\numsymant) \Bdet{I_{\numsymant}+D^{2}}^{\cohtime-\numsymant-1}}{\kappas{D^{2}}{\cohtime-\numsymant}}\sum_{k=1}^{\numsymant}\Bdet{R_{k}(I_{\numsymant}+D^{2})} 
 \label{ub_gdef_sym}
\end{IEEEeqnarray} 
\fi
where $\gammas{\cdot}$ is given in~\eqref{gammas} and $\kappas{\cdot}{\cdot}$ is given in~\eqref{kappadef}. The matrix $R_{k}(A)$ in~\eqref{ub_gdef_sym}, which is a function of the $\numsymant\times\numsymant$ diagonal matrix $A,$ is defined as follows. Let $a_1>a_2>\dots>a_{\numsymant}$ denote the ordered diagonal entries of $A$. Let the $\numsymant\times\numsymant$ real matrix $P_{k}(A)$ and the $(\cohtime-\numsymant)\times\numsymant$ real matrix $T_{k},\ k=1,2,\dots,\numsymant$ be defined as follows:
\iftwocol
\begin{IEEEeqnarray}{rCl}
 [P_{k}(A)]_{ij} &=& \left\{ \,
\begin{IEEEeqnarraybox}[][c]{l?l}
\IEEEstrut
a_i^{\numsymant-k+1}(\Blog{a_i}+\psi(\numsymant-k+1)),  & \\
\IEEEeqnarraymulticol{2}{r}{1\leq i \leq \numsymant,\  j=k} \\
a_i^{\numsymant-j+1}, &  \\
\IEEEeqnarraymulticol{2}{r}{1\leq i \leq \numsymant,\ 1\leq j\leq \numsymant,\ j\neq k}
\IEEEstrut
\end{IEEEeqnarraybox}
\right.
\end{IEEEeqnarray}
\else
\begin{IEEEeqnarray}{rCl}
 [P_{k}(A)]_{ij} &=& \left\{ \,
\begin{IEEEeqnarraybox}[][c]{l?l}
\IEEEstrut
a_i^{\numsymant-k+1}(\Blog{a_i}+\psi(\numsymant-k+1)),  & 1\leq i \leq \numsymant,\  j=k \\
a_i^{\numsymant-j+1}, &  1\leq i \leq \numsymant,\ 1\leq j\leq \numsymant,\ j\neq k
\IEEEstrut
\end{IEEEeqnarraybox}
\right.
\end{IEEEeqnarray}
\fi
and 
\iftwocol
\begin{IEEEeqnarray}{rCl}
[T_k]_{ij} &=& \left\{ \,
\begin{IEEEeqnarraybox}[][c]{ll}
\IEEEstrut
\frac{1}{\beta(\numsymant-j+1,\cohtime-\numsymant-i)}, &  \\
\IEEEeqnarraymulticol{2}{r}{\quad\quad\quad 1\leq i \leq \numsymant,\ 1\leq j\leq \numsymant,\ j\neq k} \\
\frac{\psi(\cohtime-i-k+1)}{\beta(\numsymant-k+1,\cohtime-i)}, &  \\
\IEEEeqnarraymulticol{2}{r}{1\leq i \leq \cohtime-\numsymant-1,\  j = k }\\
 1, &  \\
\IEEEeqnarraymulticol{2}{r}{i=\cohtime-\numsymant,\ 1\leq j \leq \numsymant,\ j \neq k} \\
\psi(\numsymant-k+1), & \\
\IEEEeqnarraymulticol{2}{r}{i=\cohtime-\numsymant,\ j=k}
\IEEEstrut
\end{IEEEeqnarraybox}
\right. \label{tk_def}
\end{IEEEeqnarray}
\else
\begin{IEEEeqnarray}{rCl}
[T_k]_{ij} &=& \left\{ \,
\begin{IEEEeqnarraybox}[][c]{l?l}
\IEEEstrut
\frac{1}{\beta(\numsymant-j+1,\cohtime-\numsymant-i)}, & 1\leq i \leq \cohtime-\numsymant-1,\ 1\leq j\leq \numsymant,\ j \neq k \\
\frac{\psi(\cohtime-i-k+1)}{\beta(\numsymant-k+1,\cohtime-i)}, & 1\leq i \leq \cohtime-\numsymant-1,\  j = k \\
1, &  i=\cohtime-\numsymant,\ 1\leq j \leq \numsymant,\ j \neq k \\
\psi(\numsymant-k+1), &  i=\cohtime-\numsymant,\ j=k
\IEEEstrut
\end{IEEEeqnarraybox}
\right. \label{tk_def}
\end{IEEEeqnarray}
\fi
\marktext{where~$\beta(\cdot,\cdot)$ is defined in the notation section.} Finally, let $Q(A)\in\reals^{\numsymant\times(\cohtime-\numsymant)}$ and $S\in\reals^{(\cohtime-\numsymant)\times(\cohtime-\numsymant)}$ be given by
\begin{IEEEeqnarray}{rCl}
[Q(A)]_{ij} &=& (-a_i)^{j+\numsymant-\cohtime},  \quad 1\leq i \leq \numsymant,\ 1\leq j\leq \cohtime-\numsymant \IEEEeqnarraynumspace
\end{IEEEeqnarray}
and 
\marktext{\begin{IEEEeqnarray}{rCl}
S_{ij} &=&\left\{ \,
\begin{IEEEeqnarraybox}[][c]{l?l}
\IEEEstrut
\frac{(-1)^{i-j}}{\beta(i-j+1,\cohtime-\numsymant-i)}, & 1\leq j \leq i \leq \cohtime-\numsymant-1\\
(-1)^{i-j}, & 1\leq j \leq i = \cohtime-\numsymant\\
0, &   1\leq i < j \leq \cohtime-\numsymant
\IEEEstrut .
\end{IEEEeqnarraybox}
\right. \nonumber \\ \label{smatdef}
\end{IEEEeqnarray}}
Then, 
\begin{IEEEeqnarray}{c}
R_{k}(A) = P_{k}(A)-Q(A)S^{-1}T_{k}. \label{rkdef}
\end{IEEEeqnarray}
\end{cor}
\begin{IEEEproof} See Appendix \ref{appendix_ub_symproof}.\end{IEEEproof}

In Theorem  \ref{theorem_lb} below  we provide a lower bound on $\capacity(\snr).$ This lower bound is obtained by evaluating the mutual information in~\eqref{sumratecap} for the \sphuniform input distribution introduced in Section \ref{sec_introduction}. Specifically, we set
\marktext{\begin{IEEEeqnarray}{c}
\cinp_i = \sqrt{\frac{\cohtime \snr}{\numtxant}}\bV_i
\label{spunidist}
\end{IEEEeqnarray} }with $\bV_i$ uniformly distributed on $\isotropicgrp(\ithuserant,\cohtime).$ With this choice, the power constraint in~\eqref{userpwrconst} is satisfied with equality. Similar to the bounds developed in~\cite{rusek2012mutual,hassibi2002multiple,alfano2014closed}, our lower bound relies on the Itzykson-Zuber integral~\cite[Eq. (3.4)]{itzykson1980planar}. In order to give a compact expression for our lower bound, we shall focus on the setup where each user is equipped with a single antenna i.e.,~$\ithuserant=1,\ i=1,2,\dots,\numuser.$ We will address the case when the users have multiple antennas at the end of this section.
\begin{thm} \label{theorem_lb}
The sum-rate capacity $\capacity(\snr)$ in~\eqref{sumratecap} of the Rayleigh block-fading MAC~\eqref{ysixiw} for the case when all transmitters have a single antenna, i.e., $\ithuserant=1,\ i=1,2,\dots,\numuser$ is lower-bounded as follows: 
\iftwocol
\begin{IEEEeqnarray}{rCl}
\capacity(\snr)&\geq & \Blog{\frac{\gammas{\cohtime-\numminant}}{\gammas{\cohtime}} } +\numrxant\cohtime\snr \nonumber \\ && -\>\numrxant\Bexop{\Blogdet{I_{\numtxant}+\bD^2}} \nonumber \\
&& +\>\Bexop{\Blog{\kappas{\cout\cout^{\dag}}{\cohtime-\numrxant}}} \nonumber \\
&& -\>\Bexop{\Blog{\Bexoprand{\bD}{\frac{\Bdet{M(\cout\cout^{\dag},\bE)}}{\Bdet{I_{\numtxant}+ \bD^2}^{\numrxant}\kappas{\bE}{\cohtime-\numtxant} }  }}}.  \nonumber \\ \label{exprixy}
\end{IEEEeqnarray} 
\else
\begin{IEEEeqnarray}{rCl}
\capacity(\snr) &\geq &\Blog{\frac{\gammas{\cohtime-\numminant}}{\gammas{\cohtime}} } +\numrxant\cohtime\snr   -\numrxant\Bexop{\Blogdet{I_{\numtxant}+\bD^2}} \nonumber \\
&& +\>\Bexop{\Blog{\kappas{\cout\cout^{\dag}}{\cohtime-\numrxant}}}  - \Bexop{\Blog{\Bexoprand{\bD}{\frac{\Bdet{M(\cout\cout^{\dag},\bE)}}{\Bdet{I_{\numtxant}+ \bD^2}^{\numrxant}\kappas{\bE}{\cohtime-\numtxant} }  }}}. \IEEEeqnarraynumspace \label{exprixy}
\end{IEEEeqnarray} 
\fi
Here, $\bD$ is an $\numtxant\times\numtxant$ real diagonal matrix with entries containing the singular values of the input matrix $\cinp,$ which is assumed to follow the \sphuniform distribution~\eqref{spunidist}. 
Furthermore, the~$\numtxant\times\numtxant$ matrix $\bE$ is defined as $\bE= (\bD^{-2}+I_{\numtxant})^{-1},\ \gammas{\cdot}$ is given in~\eqref{gammas}, and $\kappas{\cdot}{\cdot}$ is given in~\eqref{kappadef}. 
Finally, the $\nummaxant\times\nummaxant$ real matrix $M(A,B)$ is defined as follows:
\marktext{\iftwocol
\begin{multline}
[M(A,B)]_{ij}\\=\mathopen{}\left\{ \,
\begin{IEEEeqnarraybox}[][c]{l?l}
\IEEEstrut
   \Bupperexp{a_i b_j}{\cohtime-\nummaxant}, & 1\leq i \leq \numrxant,\ 1\leq j \leq \numtxant\\
   b_j^{\cohtime-i}, & \numrxant+1 \leq i \leq \nummaxant,\ 1\leq j \leq \numtxant \\
   a_i^{\cohtime-j}, & 1\leq i \leq \numrxant,\ \numtxant+1\leq j \leq \nummaxant.
\IEEEstrut
\end{IEEEeqnarraybox}
\right.\mathclose{}  \label{Mdef}
\end{multline} 
\else
\begin{IEEEeqnarray}{lCl}
[M(A,B)]_{ij}&=&\mathopen{}\left\{ \,
\begin{IEEEeqnarraybox}[][c]{l?l}
\IEEEstrut
   \Bupperexp{a_i b_j}{\cohtime-\nummaxant}, & 1\leq i \leq \numrxant,\ 1\leq j \leq \numtxant\\
   b_j^{\cohtime-i}, & \numrxant+1 \leq i \leq \nummaxant,\ 1\leq j \leq \numtxant \\
   a_i^{\cohtime-j}, & 1\leq i \leq \numrxant,\ \numtxant+1\leq j \leq \nummaxant.
\IEEEstrut
\end{IEEEeqnarraybox}
\right.\mathclose{}  \label{Mdef}
\end{IEEEeqnarray} 
\fi }
Here, $a_1>a_2>\dots>a_{\numrxant}$ are the ordered eigenvalues of the positive-definite matrix~$A\in\complexset^{\numrxant\times\numrxant}$; similarly $b_1>b_2>\dots>\marktext{b_{\numtxant}}$ are the ordered eigenvalues of the positive-definite matrix~$B\in\complexset^{\numtxant\times\numtxant};$ the function $\Bupperexp{\cdot}{\cdot}$ in~\eqref{Mdef} is given in~\eqref{gamma_x_n_def}.
\end{thm}
\begin{IEEEproof} See Appendix \ref{appendix_lb_proof}.
\end{IEEEproof}

The lower bound~\eqref{exprixy} involves expectations that are not known in closed form.
Hence, we resort to Monte-Carlo methods for the evaluation of~\eqref{exprixy}. 
One exception is the two-user case $\numuser=2,$ for which the expectation over $\bD$ in~\eqref{exprixy} admits a closed-form integral expression. This result is presented in Corollary \ref{cor_lb_2x2} below. The proof of this corollary exploits that for the two-user case, the probability distribution of the eigenvalues of $\cinp\cinp^{\dag}$ can be obtained from the p.d.f. of $\Bdet{\cinp\cinp^{\dag}},$ which can be calculated through Bartlett's decomposition~\cite[Prop. 2.1]{rouault2006asymptotic}. This approach does not extend to the case $\numuser>2$ or $\numtxant>2.$
\begin{cor} \label{cor_lb_2x2}
The sum-rate capacity $\capacity(\snr)$ in~\eqref{sumratecap} of the Rayleigh block-fading MAC~\eqref{ysixiw} with~$\numuser=2$ and $\ithuserant=1,\ i=1,2$ is lower-bounded as follows:
\marktext{\iftwocol
\begin{IEEEeqnarray}{rCl}
\IEEEeqnarraymulticol{3}{l}{\capacity(\snr) \geq \Blog{\frac{\gammas{\cohtime-\numminant}}{\gammas{\cohtime}} } +\numrxant\cohtime\snr+\Blog{\frac{\cohtime\snr}{2}}}\nonumber\\
&&-\>\numrxant(\cohtime-1)\left[\int_0^1\Blog{\mu^2-\alpha}(1-\alpha)^{\cohtime-2}d\alpha\right]\nonumber \\
&& +\> \Bexop{\Blog{\kappas{\cout\cout^{\dag}}{\cohtime-\numrxant}}} - \Blog{\cohtime-1}\nonumber \\
&& -\> \Bexop{\Blog{\int_0^1 \Bdet{M(\cout\cout^{\dag},E(\alpha))} \left(\mu^2-\alpha^2\right)^{\cohtime-\numrxant-1}d\alpha}}. \nonumber \\  \label{exprixy2case}
\end{IEEEeqnarray}
\else
\begin{IEEEeqnarray}{rCl}
\capacity(\snr) &\geq &\Blog{\frac{\gammas{\cohtime-\numminant}}{\gammas{\cohtime}} } +\numrxant\cohtime\snr+\Blog{\frac{\cohtime\snr}{2}}-\numrxant(\cohtime-1)\left[\int_0^1\Blog{\mu^2-\alpha}(1-\alpha)^{\cohtime-2}d\alpha\right]\nonumber \\
&& +\> \Bexop{\Blog{\kappas{\cout\cout^{\dag}}{\cohtime-\numrxant}}} - \Blog{\cohtime-1}\nonumber \\
&& -\> \Bexop{\Blog{\int_0^1 \Bdet{M(\cout\cout^{\dag},E(\alpha))} \left(\mu^2-\alpha^2\right)^{\cohtime-\numrxant-1}d\alpha}}.  \label{exprixy2case}
\end{IEEEeqnarray} 
\fi }
Here, $\mu = 1+2/\cohtime\snr$ and 
\begin{IEEEeqnarray}{rCl}
E(\alpha)&=&\DMat{\frac{1+\alpha}{1+\mu(1+\alpha)},\frac{1-\alpha}{1+\mu(1-\alpha)}}.
\end{IEEEeqnarray} Furthermore, $M(\cdot,\cdot)$ is given in~\eqref{Mdef}, $\gammas{\cdot}$ is given in~\eqref{gammas}, and $\kappas{\cdot}{\cdot}$ is given in~\eqref{kappadef}.
\end{cor}
\begin{IEEEproof}
See Appendix \ref{appendix_lb_2x2_proof}.
\end{IEEEproof}

For the case when the users are equipped with multiple antennas, the singular values of the input matrix $\cinp$ are no longer distinct, and~\eqref{exprixy} needs to be further simplified using  L'Hôpital's rule. The final expression of the resulting lower bound is omitted because it is involved. Instead, numerical results are provided in Section \ref{sec_sim_results}.

\section{Numerical Results}
\label{sec_sim_results}
Focusing on the setup where the total number of transmit antennas $\numtxant$ is equal to the number of receive antennas $\numrxant,$ we numerically evaluate in this section the upper bound~\eqref{expr_ub_sym} and the lower bound~\eqref{exprixy} on the sum-rate capacity~\eqref{sumratecap}.\footnote{Numerical routines implementing the upper bound~\eqref{expr_ub_sym} and lower bound~\eqref{exprixy} can be downloaded at \url{https://github.com/infotheorychalmers/mac_capacity_bounds}}

We also evaluate the \iid Gaussian lower bound developed by Rusek \etal~\cite{rusek2012mutual}, which is obtained by setting $\cinp = \sqrt{\snr/\numtxant}\bG$ in~\eqref{sumratecap}, with $\bG$ having \iid $\cnormdist{0}{1}$ entries. For this choice of $\cinp,$ the expectation over $\bD$ in~\eqref{exprixy} can be evaluated in closed form using the properties of Wishart matrices and the integral formula~\cite[Cor. 2]{chiani2003capacity}. See~\cite{rusek2012mutual,alfano2014closed} for details.

We also consider the simple capacity upper bound obtained by assuming that the receiver has perfect knowledge of the channel matrix $\channel$ in~\eqref{sdef}. This bound is given by~\cite{teletar1999CoherentCapacity} \begin{IEEEeqnarray}{rCl}
\capacity(\snr) &\leq &\Bexop{ \Blogdet{I_{\numrxant}+\frac{\snr}{\numtxant} \channel\channel^{\dag}}}. \label{cohcap_ub}
\end{IEEEeqnarray}

\begin{figure}
\centering
\includegraphics{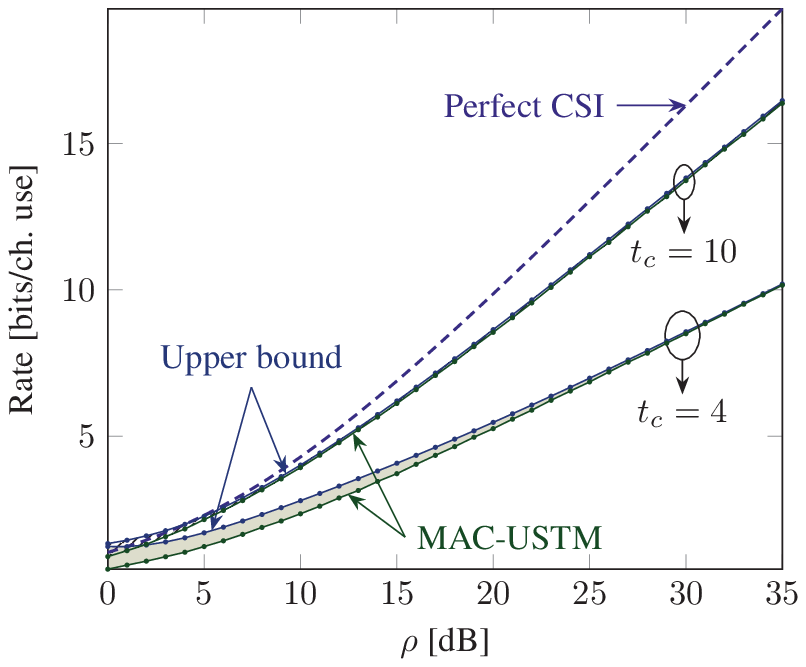}
\caption{Upper bounds~\eqref{expr_ub_sym} and~\eqref{cohcap_ub}, and lower bound~\eqref{exprixy} on the sum-rate capacity of the MAC~\eqref{ysixiw}; $\numuser=2$ single-antenna users, $\numrxant=2,\ \cohtime\in\{4,10\}.$}
\label{fig_sim_results2x2}
\end{figure}

\begin{figure}
\centering
\includegraphics{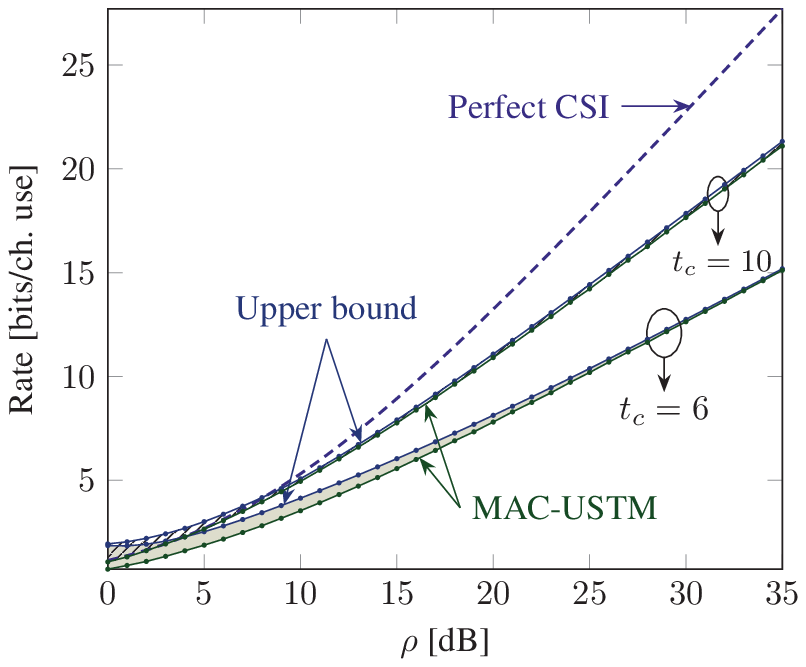}
\caption{Upper bounds~\eqref{expr_ub_sym} and~\eqref{cohcap_ub}, and lower bound~\eqref{exprixy} on the sum-rate capacity of the MAC~\eqref{ysixiw}; $\numuser=3$ single-antenna users, $\numrxant=3,\ \cohtime\in\{6,10\}.$}
\label{fig_sim_results3x3}
\end{figure}

\begin{figure}
\centering
\includegraphics{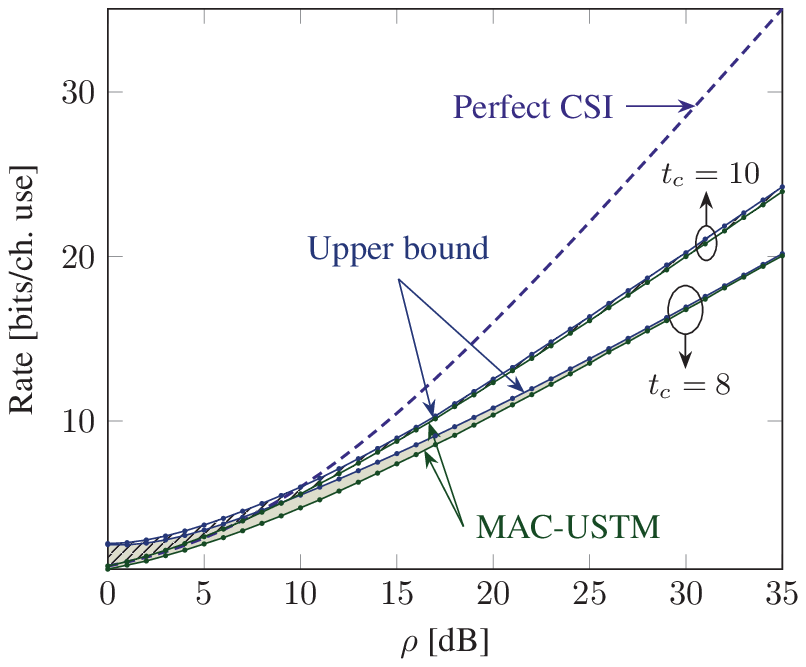}
\caption{Upper bounds~\eqref{expr_ub_sym} and~\eqref{cohcap_ub}, and lower bound~\eqref{exprixy} on the sum-rate capacity of the MAC~\eqref{ysixiw}; $\numuser=4$ single-antenna users, $\numrxant=4,\ \cohtime\in\{8,10\}.$}
\label{fig_sim_results4x4}
\end{figure}

We start by considering the case of single-antenna users. In Figs.~\cref{fig_sim_results2x2,fig_sim_results3x3,fig_sim_results4x4}, we depict the upper bounds~\eqref{expr_ub_sym} and~\eqref{cohcap_ub}, and the lower bound~\eqref{exprixy} on the sum-rate capacity of the MAC~\eqref{ysixiw} for different values of number of users $\numuser,$  number of receive antennas $\numrxant,$ and  coherence interval~$\cohtime$. 
For the choice of parameters in Figs.~\cref{fig_sim_results2x2,fig_sim_results3x3,fig_sim_results4x4}, the upper bound~\eqref{expr_ub_sym} and the lower bound~\eqref{exprixy} characterize the sum-rate capacity accurately already at SNR values as low as $10$~dB. The tightness of the bounds increases as the coherence interval $\cohtime$ or the SNR increases. This last observation comes as no surprise, since the choice of the auxiliary distribution in the  derivation of the upper bound~\eqref{expr_ub_sym} is dictated by high-SNR considerations.

The tightness of our bounds implies that the sum-rate capacity of the MAC~\eqref{ysixiw} is well-approximated by the capacity of an equivalent point-to-point MIMO channel with the same SNR and the same total number of transmit and receive antennas. This follows because the upper bound is derived under the assumption of perfect cooperation among the users. Furthermore, we observe that the perfect-CSI sum-rate capacity upper bound~\eqref{cohcap_ub} is loose for the relatively small $\cohtime$ values considered in this section.

As shown in Fig.~\ref{fig_intro_cmp} (see Section \ref{sec_introduction}) for the case $\numuser=\numrxant=4$ and $\cohtime=10,$ the \sphuniform lower bound is tighter than the \iid Gaussian lower bound in the high-SNR regime, although marginally so. The same consideration holds for the case $\numuser=\numrxant=2$ and $\numuser=\numrxant=3$ considered in Fig.~\ref{fig_sim_results2x2} and Fig.~\ref{fig_sim_results3x3}. Furthermore, the gap between the two bounds gets smaller for~$\cohtime$ values smaller than $10.$

\begin{figure}
\centering
\includegraphics{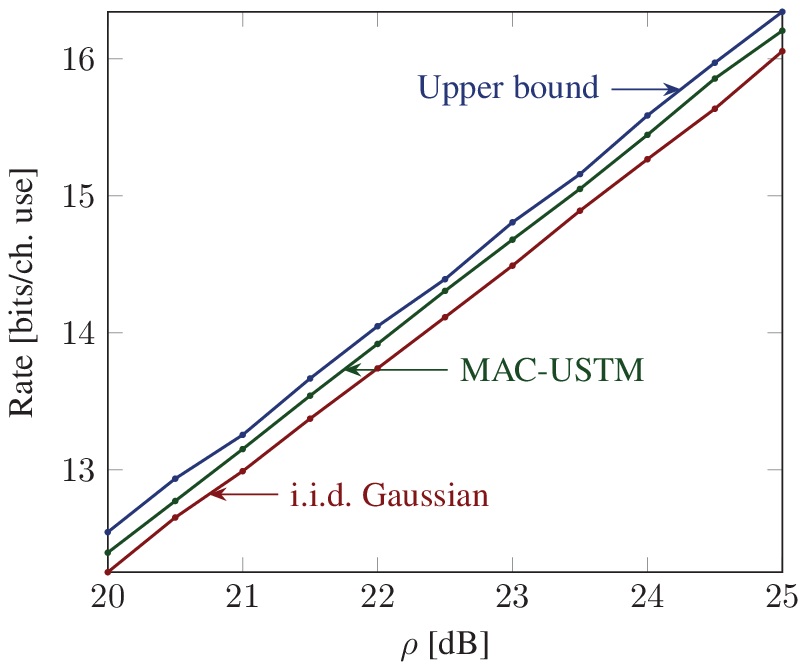}
\caption{Upper bound~\eqref{expr_ub_sym}, MAC-USTM lower bound~\eqref{exprixy}, and \iid Gaussian lower bound~\cite{rusek2012mutual}; $\numuser=2$ with $n_1=1$ and $n_2=3;$ furthermore, $\numrxant=4$ and $\cohtime=10.$}
\label{fig_sim_results4x4_moreusers}
\end{figure}

In Fig.~\ref{fig_sim_results4x4_moreusers}, we consider the case when $\numtxant=\numrxant=4$ and $\cohtime=10,$ but the available $4$ antennas are divided unevenly among the users. Specifically, we assume $\numuser=2$ (two users) and that the first user has one antenna $(n_1=1)$ whereas the second user has three antennas $(n_2=3).$ Note that the upper bound and the \iid Gaussian lower bound depend only on the total number of antennas $\numtxant$ and not on the way the antennas are divided among users. Hence, these two curves coincide with the curves drawn in Fig.~\ref{fig_intro_cap} and Fig.~\ref{fig_intro_cmp}. On the contrary, the \sphuniform bound depends on $n_1$ and $n_2,$ and, for the asymmetric antenna scenario considered in Fig.~\ref{fig_sim_results4x4_moreusers}, is tighter than the \iid Gaussian bound, although  marginally so. 

\section{Conclusion}
\label{sec_conclusion}

We presented finite-SNR upper and lower bounds on the sum-rate capacity of Rayleigh block-fading channels for the scenario where neither the transmitters nor the receiver have access to \apriori CSI. The upper bound, which is based on duality, is derived under the assumption that the transmitters can cooperate perfectly. This transforms the MAC into an equivalent point-to-point MIMO channel. We obtain the lower bound by assuming that a USTM signal is independently transmitted by each user over its available antennas (\sphuniform). The gap between the upper and the lower bounds is less than $8\%$ at $10$~dB for the case of $4$ single-antenna transmitters communicating with a $4$-antenna receiver over a MAC with coherence interval $10.$ The gap reduces further when the SNR or the coherence interval get larger. This implies that the capacity gains obtainable by allowing cooperation among transmitters are minimal \marktext{in this case}. Our numerical results show also that \sphuniform yields rates that are only marginally larger than the ones obtainable using \iid Gaussian inputs. This suggests that \iid Gaussian inputs are almost sum-rate capacity optimal for the Rayleigh block-fading MACs considered in Section~\ref{sec_sim_results}.

\begin{appendices}
\section{Preliminary Results}
\label{appendix_pre_result}
\subsection{An Integral Formula}
The following lemma is useful in the evaluation of integrals that involve matrix determinants. We shall need this lemma in the proof of Corollary~\ref{theorem_ub_sym}.
\begin{lem}
\label{lemmaintegralformula}
Let $a,b\in\reals,$ with $a<b.$ Let $\listofvars{f_i(\cdot)}{i=1}{n}$ and $\listofvars{g_i(\cdot)}{i=1}{n}$ be arbitrary integrable functions over $[a,b).$ Let $A=A(x_1,x_2,\dots,x_n)$ and $B=B(x_1,x_2,\dots,x_n)$ be $n\times n$ and $m\times m$ matrices $(m\geq n)$ whose entries depend on the scalars $x_1,x_2,\dots,x_n$ as follows:
\begin{IEEEeqnarray}{rCl}
A_{ij} &=&\left\{ \,
\begin{IEEEeqnarraybox}[][c]{l?l}
\IEEEstrut
f_{i}(x_j), & 1\leq i \leq m,\ 1 \leq j \leq n \\
c_{ij}, & 1\leq i \leq m,\ n+1 \leq j \leq m 
\IEEEstrut
\end{IEEEeqnarraybox}
\right. \\
B_{ij} &=& g_{i}(x_j), \quad 1\leq i \leq n,\ 1 \leq j \leq n
\end{IEEEeqnarray}
where $c_{ij}$ are arbitrary scalar real constants.
Finally, let $E\in\reals^{m\times m}$ be defined as
\begin{IEEEeqnarray}{rCl}
E_{ij} &=& \left\{ \,
\begin{IEEEeqnarraybox}[][c]{l?l}
\IEEEstrut
\int_{a}^{b} f_{i}(x)g_{j}(x)dx, & 1\leq i \leq m,\ 1 \leq j \leq n \\
c_{ij}, & 1\leq i \leq m,\ n+1 \leq j \leq m.
\IEEEstrut
\end{IEEEeqnarraybox}
\right. 
\end{IEEEeqnarray} 
Then
\iftwocol
\begin{IEEEeqnarray}{c}
 \underset{{a\leq x_{n}\leq x_{n-1} \dots \leq x_1 < b} }{\idotsint}\Bdet{A}\Bdet{B}dx_1dx_2\dots dx_n = \Bdet{E}. \nonumber \\ \label{expr_integralformula}
\end{IEEEeqnarray}
\else
\begin{IEEEeqnarray}{c}
 \underset{{a\leq x_{n}\leq x_{n-1} \dots \leq x_1 < b} }{\idotsint}\Bdet{A}\Bdet{B}dx_1dx_2\dots dx_n = \Bdet{E}.  \label{expr_integralformula}
\end{IEEEeqnarray}
\fi
\end{lem}
\begin{IEEEproof}See~\cite[Lem. 2]{1687741}. \end{IEEEproof}

\subsection{Limits of Determinants}
The following lemma, which characterizes the limiting behavior of the ratio between the determinant of a certain matrix and a Vandermonde determinant, will be needed in the proof of Corollary~\ref{theorem_ub_sym}. 

\begin{lem} \label{lemmalhoprule}
Let $A\in \complexset^{m\times m}$ be a positive-definite matrix with ordered eigenvalues $a_1>a_2>\dots>a_m.$ Let $ C \in \complexset^{m\times m} $ be a matrix with entries  $C_{i,j} = f_i(a_j),$ for some differentiable functions~$\listofvars{f_i(\cdot)}{i=1}{m}.$ \marktext{Then for every integer~$n<m,$ and every scalar real constant~$a_0,$}
\iftwocol
\begin{multline}
\lim_{\{a_{n+1},a_{n+2},\dots,a_{m}\}\rightarrow a_0}\frac{\Bdet{C}}{\vand(A)}\\ = \frac{\Bdet{E}}{\gammas{m-n}\kappas{A_0-a_0I_n}{m-n} } \label{expr_lemmalhoprule}
\end{multline}
\else
\begin{IEEEeqnarray}{rCl}
\lim_{\{a_{n+1},a_{n+2},\dots,a_{m}\}\rightarrow a_0}\frac{\Bdet{C}}{\vand(A)} &=& \frac{\Bdet{E}}{\gammas{m-n}\kappas{A_0-a_0I_n}{m-n} } \label{expr_lemmalhoprule}
\end{IEEEeqnarray}
\fi
where  $A_0 = \DMat{a_1,a_2,\dots,a_n}, \gammas{\cdot}$ is given in~\eqref{gammas}, $\vand(\cdot)$ is given in~\eqref{vanderdet}, $\kappas{\cdot}{\cdot}$ is given in~\eqref{kappadef},  and the entries of the $m\times m$ matrix $E$ are \begin{IEEEeqnarray}{rCl}
E_{i,j} &= \left\{ \,
\begin{IEEEeqnarraybox}[][c]{l?l}
\IEEEstrut
f_i(a_j), & 1\leq i \leq m,\ 1\leq j \leq n\\
f_i^{(m-j)}(a_0), & 1 \leq i \leq m,\ n+1\leq j \leq m.
\IEEEstrut
\end{IEEEeqnarraybox}
\right.
\end{IEEEeqnarray} Here, $f_i^{(k)}(\cdot)$ denotes the $k$th-order derivative of $f_i(\cdot).$
\end{lem}
\begin{IEEEproof} See~\cite[Lem. 5]{6185738}. \end{IEEEproof}

\subsection{Expectation of the Log Determinant of a Gaussian Quadratic Form} \label{appendixlogdet}
The following lemma gives a closed-form expression for $\Bexop{\Blogdet{\bX L\bX^{\dag}}}$ where $\bX$ has \iid $\cnormdist{0}{1}$ entries and $L$ is a certain positive-definite matrix. A closed-form expression for the case when the eigenvalues of $L$ are distinct is provided in~\cite[Lem. 2]{1542408}. Here, we derive a different closed-form expression, which does not require the eigenvalues to be distinct, and appears better suited for numerical evaluations. This lemma is used in the proof of Corollary~\ref{theorem_ub_sym}.

\begin{lem} \label{lem_exp_log_det_gauss}
Let $\bX\in \complexset^{n\times m},$ where $n<m,$ be a random matrix with \iid $\cnormdist{0}{1}$ entries. Let $L\in\complexset^{m\times m}$ be a positive-definite matrix whose largest $n$ eigenvalues satisfy $l_1>l_2>\dots>l_n>1,$ and whose lowest $m-n$ eigenvalues are equal to 1. Then  
\iftwocol
\begin{multline}
\Bexop{\Blogdet{\bX L\bX^{\dag}}} \\= \frac{ \Bdet{L_0}^{m-n-1}}{\kappas{L_0-I_n}{m-n}}  \sum_{k=1}^{n}\Bdet{R_{k}(L_0)} \label{expr_exp_log_det}
\end{multline} 
\else
\begin{IEEEeqnarray}{rCl}
\Bexop{\Blogdet{\bX L\bX^{\dag}}} = \frac{ \Bdet{L_0}^{m-n-1}}{\kappas{L_0-I_n}{m-n}}  \sum_{k=1}^{n}\Bdet{R_{k}(L_0)} \label{expr_exp_log_det}
\end{IEEEeqnarray} 
\fi
where $\kappas{\cdot}{\cdot}$ is given in~\eqref{kappadef}, $L_0=\DMat{l_1,l_2,\dots,l_n}$, and $R_{k}(\cdot)$ is defined as in Corollary~\ref{theorem_ub_sym}.
\end{lem}
\begin{IEEEproof}
We first obtain the joint p.d.f. of the eigenvalues of $\bS^{*} = \bX L^{*}\bX^{\dag},$ where $L^{*}\in\complexset^{m\times m}$ is a positive-definite matrix whose largest $n$ eigenvalues coincide with the eigenvalues of~$L,$ and whose remaining $m-n$ eigenvalues are distinct, i.e., $l_{n+1}^{*}>l_{n+2}^{*}>\dots>l_{m}^{*}.$ We then derive the p.d.f. of the eigenvalues of $\bS = \bX L\bX^{\dag}$ by letting $\{l_{n+1}^{*},l_{n+2}^{*},\dots,l_{m}^{*}\}\rightarrow 1.$ Knowledge of this p.d.f. allows us to obtain the moment generating function of $\Blogdet{\bS}$ in closed form, from which~\eqref{expr_exp_log_det} follows by evaluating the derivative of this moment generating function at zero. 
The p.d.f. of $\bS^{*}$ is~\cite[Sec. 2]{gao2000determinant}\footnote{There seems to be a typo in the expression given in~\cite[Sec. 2]{gao2000determinant}. 
The term  $\pi^{n(n-1)/2}$ should be in denominator (see~\cite[Eq. (50) and Eq. (57)]{khatri1966certain}).}
\iftwocol
\begin{multline}
f_{\bS^{*}}(S)\\ = \frac{1}{\pi^{n(n-1)/2}}\frac{ \Bdet{F(-(L^{*})^{-1},S)}}{\vand(S)\vand(-(L^{*})^{-1})\Bdet{{L^{*}}}^{n} },\quad S\in \complexset^{n\times n}
\end{multline} 
\else
\begin{IEEEeqnarray}{c}
f_{\bS^{*}}(S) = \frac{1}{\pi^{n(n-1)/2}}\frac{ \Bdet{F(-(L^{*})^{-1},S)}}{\vand(S)\vand(-(L^{*})^{-1})\Bdet{{L^{*}}}^{n} },\quad S\in \complexset^{n\times n}
\end{IEEEeqnarray}  
\fi
where $F(A,B)$ is an $m\times m$ matrix that depends on the ordered eigenvalues $\listofvars{a_i}{i=1}{m}$ and $\listofvars{b_i}{i=1}{n}$ of the Hermitian matrices $A\in \complexset^{m\times m}$ and $B\in \complexset^{n\times n}$ as follows
\marktext{\iftwocol
\begin{IEEEeqnarray}{rCl}
[F(A,B)]_{ij} &=& \left\{ \,
\begin{IEEEeqnarraybox}[][c]{l?l}
\IEEEstrut
e^{a_ib_j}, & 1\leq i \leq m,\ 1\leq j\leq n \\
a_i^{m-j},& 1\leq i \leq m,\ n+1\leq j\leq m. 
\IEEEstrut
\end{IEEEeqnarraybox}
\right. \IEEEeqnarraynumspace \label{expr_f_a_b}
\end{IEEEeqnarray} 
\else
\begin{IEEEeqnarray}{rCl}
[F(A,B)]_{ij} &=& \left\{ \,
\begin{IEEEeqnarraybox}[][c]{l?l}
\IEEEstrut
\Bexp{a_ib_j}, & 1\leq i \leq m,\ 1\leq j\leq n \\
a_i^{m-j},& 1\leq i \leq m,\ n+1\leq j\leq m. 
\IEEEstrut
\end{IEEEeqnarraybox}
\right. \label{expr_f_a_b}
\end{IEEEeqnarray} 
\fi }
Let $\bE_{\bS^{*}}$ denote the $n\times n$ real diagonal matrix having the ordered eigenvalues of $\bS^{*}$ on its main diagonal. The p.d.f. of $\bE_{\bS^{*}}$ is~\cite[Thm. 3.2]{edelman1989eigenvalues}
\iftwocol
\begin{multline}
f_{\bE_{\bS^{*}}}(E)\\= \frac{1}{\gammas{n}} \frac{ \Bdet{F(-(L^{*})^{-1},E)}\vand(E)}{\vand(-(L^{*})^{-1})\Bdet{{L^{*}}}^{n} },\quad E\in\diagmatset{n}{n}. \label{expr_appendix_e_s_star_dist}
\end{multline} 
\else
\begin{IEEEeqnarray}{c}
f_{\bE_{\bS^{*}}}(E)= \frac{1}{\gammas{n}} \frac{ \Bdet{F(-(L^{*})^{-1},E)}\vand(E)}{\vand(-(L^{*})^{-1})\Bdet{{L^{*}}}^{n} },\quad E\in\diagmatset{n}{n}. \label{expr_appendix_e_s_star_dist}
\end{IEEEeqnarray} 
\fi
Finally, let $\bE_{\bS}$ denote the $n\times n$ real diagonal matrix having the  ordered eigenvalues of $\bS=\bX L\bX^{\dag}$ on its main diagonal. 
We  obtain the p.d.f. of $\bE_{\bS}$ by computing the limit $\{l_{n+1}^{*},l_{n+2}^{*},\dots,l_{m}^{*}\}\rightarrow 1$ in~\eqref{expr_appendix_e_s_star_dist}:
\iftwocol
\begin{IEEEeqnarray}{rCl}
\IEEEeqnarraymulticol{3}{l}{f_{\bE_{\bS}}(E)=  \lim_{\{l_{n+1}^{*},l_{n+2}^{*},\dots,l_{m}^{*}\}\rightarrow 1}f_{\bE_{\bS^{*}}}(E)}\\
&=&\frac{\vand(E)}{\gammas{n}\Bdet{L_0}^{n}} \frac{ \Bdet{G(L_0,E)}}{\gammas{m-n}\kappas{-L_0^{-1}+I_n}{m-n}} \label{expr_appendix_e_s_dist1}\\
&=&\frac{ \Bdet{L_0}^{m-n-1} \Bdet{G(L_0,E)}\vand(E)}{\gammas{n}\gammas{m-n}\kappas{L_0-I_n}{m-n}},\quad E \in \diagmatset{n}{n}.\IEEEeqnarraynumspace \label{expr_appendix_e_s_dist_final}
\end{IEEEeqnarray}
\else
\begin{IEEEeqnarray}{rCl}
f_{\bE_{\bS}}(E) &= & \lim_{\{l_{n+1}^{*},l_{n+2}^{*},\dots,l_{m}^{*}\}\rightarrow 1}f_{\bE_{\bS^{*}}}(E)\\
&=&\frac{\vand(E)}{\gammas{n}\Bdet{L_0}^{n}} \frac{ \Bdet{G(L_0,E)}}{\gammas{m-n}\kappas{-L_0^{-1}+I_n}{m-n}} \label{expr_appendix_e_s_dist1}\\
&=&\frac{ \Bdet{L_0}^{m-n-1} \Bdet{G(L_0,E)}\vand(E)}{\gammas{n}\gammas{m-n}\kappas{L_0-I_n}{m-n}},\quad E \in \diagmatset{n}{n}. \label{expr_appendix_e_s_dist_final}
\end{IEEEeqnarray}
\fi
Here, $L_0\in\diagmatset{n}{n}$ is a diagonal matrix, which contains the $n$ largest eigenvalues $l_1>l_2>\dots>l_n$ of $L$ on its diagonal. Furthermore, the $m\times m$ matrix $G(L_0,E)$ is defined as follows
\iftwocol
\begin{IEEEeqnarray}{l}
[G(L_0,E)]_{ij} =\nonumber\\ \left\{ \,
\begin{IEEEeqnarraybox}[][c]{l?l}
\IEEEstrut
\Bexp{-l_i^{-1}e_j}, & 1\leq i \leq n,\ 1\leq j\leq n\\
e_j^{m-i}\Bexp{-e_j}, & n+1\leq i \leq m,\ 1\leq j\leq n\\
(-l_i)^{j-m}, & 1\leq i \leq n,\ n+1\leq j\leq m \IEEEeqnarraynumspace\\
\frac{\Gamma(m-j+1)}{\Gamma(i-j+1)}(-1)^{i-j}, & n+1\leq j \leq i \leq m \\
0, & n+1\leq i < j \leq m
\IEEEstrut
\end{IEEEeqnarraybox}
\right.
\end{IEEEeqnarray}
\else
\begin{IEEEeqnarray}{l}
[G(L_0,E)]_{ij} = \left\{ \,
\begin{IEEEeqnarraybox}[][c]{l?l}
\IEEEstrut
\Bexp{-l_i^{-1}e_j}, & 1\leq i \leq n,\ 1\leq j\leq n\\
e_j^{m-i}\Bexp{-e_j}, & n+1\leq i \leq m,\ 1\leq j\leq n\\
(-l_i)^{j-m}, & 1\leq i \leq n,\ n+1\leq j\leq m \IEEEeqnarraynumspace\\
\frac{\Gamma(m-j+1)}{\Gamma(i-j+1)}(-1)^{i-j}, & n+1\leq j \leq i \leq m \\
0, & n+1\leq i < j \leq m
\IEEEstrut
\end{IEEEeqnarraybox}
\right.
\end{IEEEeqnarray}
\fi
where $e_1>e_2>\cdots> e_n$ denote the diagonal entries of $E.$ To obtain~\eqref{expr_appendix_e_s_dist1}, we used Lemma~\ref{lemmalhoprule}, and~\eqref{expr_appendix_e_s_dist_final} follows because for all $A\in \complexset^{n\times n},$ we have that $\vand(-A) = \vand(A)$ and that $\vand(A^{-1})=\vand(A)/\Bdet{A}^{n-1}.$ 

Next, we evaluate the moment generating function of $\Blogdet{\bS}$:  \begin{IEEEeqnarray}{rCl}
g(t)&=& \Bexop{\Bexp{t\Blogdet{\bS}}} =  \Bexop{\Bdet{\bE_\bS}^t}\\
&=& \underset{\diagmatset{n}{n} }{\idotsint}\Bdet{E}^t f_{\bE_\bS}(E) dE \\
&=& \frac{ \Bdet{L_0}^{m-n-1} \Bdet{H(t,L_0)}}{\gammas{n}\gammas{m-n}\kappas{L_0-I_n}{m-n}}. \label{intformuse}
\end{IEEEeqnarray} Here, the $m\times m$ matrix $H(t,L_0)$ is defined as follows: 
\iftwocol
\begin{IEEEeqnarray}{l}
[H(t,L_0)]_{ij} =\nonumber \\
\left\{ \,
\begin{IEEEeqnarraybox}[][c]{l?l}
\IEEEstrut
\Gamma(t+n-j+1) l_i^{t+n-j+1}, & 1\leq i \leq n,\ 1\leq j\leq n\\
\Gamma(t+m-i+n-j+1), & n+1\leq i \leq m,\ 1\leq j\leq n \\
(-l_i)^{j-m}, & 1\leq i \leq n,\ n+1\leq j\leq m \\
\frac{\Gamma(m-j+1)}{\Gamma(i-j+1)}(-1)^{i-j}, &  n+1\leq j \leq i \leq m \\
0, &  n+1\leq i < j \leq m.
\IEEEstrut 
\end{IEEEeqnarraybox}
\right.  \nonumber \\
\end{IEEEeqnarray} 
\else
\begin{IEEEeqnarray}{l}
[H(t,L_0)]_{ij} =
\left\{ \,
\begin{IEEEeqnarraybox}[][c]{l?l}
\IEEEstrut
\Gamma(t+n-j+1) l_i^{t+n-j+1}, & 1\leq i \leq n,\ 1\leq j\leq n\\
\Gamma(t+m-i+n-j+1), & n+1\leq i \leq m,\ 1\leq j\leq n \\
(-l_i)^{j-m}, & 1\leq i \leq n,\ n+1\leq j\leq m \\
\frac{\Gamma(m-j+1)}{\Gamma(i-j+1)}(-1)^{i-j}, &  n+1\leq j \leq i \leq m \\
0, &  n+1\leq i < j \leq m.
\IEEEstrut 
\end{IEEEeqnarraybox}
\right.  
\end{IEEEeqnarray} 
\fi
To obtain~\eqref{intformuse}, we used the integral formula~\eqref{expr_integralformula} in Lemma \ref{lemmaintegralformula}. To establish~\eqref{expr_exp_log_det}, we now proceed as follows. Let $g'(t)$ denote the first derivative of $g(t).$ Then 
\iftwocol
\begin{IEEEeqnarray}{rCl}
\Bexop{\Blogdet{\bY}}&=&\lim_{t\rightarrow 0}g'(t)\\
 &=&  \frac{ \Bdet{L_0}^{m-n-1}\sum_{k=1}^{n}\Bdet{H_k(L_0)}}{\gammas{n}\gammas{m-n}\kappas{L_0-I_n}{m-n}}   \label{exprnhk} \IEEEeqnarraynumspace
\end{IEEEeqnarray} 
\else
\begin{IEEEeqnarray}{rCl}
\Bexop{\Blogdet{\bY}}&=&\lim_{t\rightarrow 0}g'(t)\\
&=& \frac{ \Bdet{L_0}^{m-n-1}\sum_{k=1}^{n}\Bdet{H_k(L_0)}}{\gammas{n}\gammas{m-n}\kappas{L_0-I_n}{m-n}}   \label{exprnhk}
\end{IEEEeqnarray} 
\fi
where the $m\times m$ matrices $H_k(L_0),\ k=1,2,\dots,n,$ are given by
\iftwocol
\begin{IEEEeqnarray}{l}
[H_k(L_0)]_{ij} =\nonumber\\
\left\{ \,
\begin{IEEEeqnarraybox}[][c]{l?l}
\IEEEstrut
{[H(0,L_0)]}_{ij}, &\\
\IEEEeqnarraymulticol{2}{r}{1\leq i \leq m,\ 1\leq j \leq m,\ j\neq k}\\
\Gamma(n-k+1) l_i^{n-k+1}(\Blog{l_i}+\psi(n-k+1)), & \\
\IEEEeqnarraymulticol{2}{r}{ 1\leq i \leq n,\ j=k}\\
\Gamma(m-i+n-k+1)\psi(m-i+n-k+1), & \\
 \IEEEeqnarraymulticol{2}{r}{n+1 \leq i \leq m,\ j=k}.
\IEEEstrut
\end{IEEEeqnarraybox}
\right.  \label{hkdef}
\end{IEEEeqnarray} 
\else
\begin{IEEEeqnarray}{rCl}
[H_k(L_0)]_{ij} =\left\{ \,
\begin{IEEEeqnarraybox}[][c]{l?l}
\IEEEstrut
{[H(0,L_0)]}_{ij}, & 1\leq i \leq m,\ 1\leq j \leq m,\ j\neq k\\
\Gamma(n-k+1) l_i^{n-k+1}(\Blog{l_i}+\psi(n-k+1)), & 1\leq i \leq n,\ j=k\\
\Gamma(m-i+n-k+1)\psi(m-i+n-k+1), & n+1 \leq i \leq m,\ j=k.
\IEEEstrut
\end{IEEEeqnarraybox}
\right.  \label{hkdef}
\end{IEEEeqnarray} 
\fi
In~\eqref{exprnhk}, we have written the derivative of the determinant of the matrix $H(t,L_0)$ as a sum of~$n$ determinants by using that the matrix $H(t,L_0)$ depends on $t$ only through its first $n$ columns.
Through algebraic manipulation one can show that
\begin{IEEEeqnarray}{rCl}
\Bdet{H_k(L_0)} &=& \gammas{n}\gammas{m-n}\Bdet{R_k(L_0)} \label{partitionhk}
\end{IEEEeqnarray}
where $R_k(\cdot)$ is defined in~\eqref{rkdef}. The equality in~\eqref{partitionhk} relates the determinant of an $m\times m$ matrix to the determinant of a smaller $n\times n$ matrix. Substituting~\eqref{partitionhk} into~\eqref{exprnhk} we obtain~\eqref{expr_exp_log_det}.
\end{IEEEproof}

\section{ Proof of Theorem~\ref{theorem_ub}}
\label{appendix_ubproof}
We upper-bound the sum-rate capacity $\capacity(\snr)$ by enlarging the set over which the supremum in~\eqref{sumratecap} is computed. Specifically, we drop the assumption that the $\listofvars{\cinp_i}{i=1}{\numuser}$ are independent, and we substitute~\eqref{userpwrconst} with the ``global" power constraint
\begin{IEEEeqnarray}{c}
\Bexop{\Btrace{\cinp\cinp^{\dag}}} \leq \cohtime\snr.\label{mimo_pwr_const}\end{IEEEeqnarray}
Let $\mimoinpdistclass(\snr)$ denote this enlarged set. 
Then
\begin{IEEEeqnarray}{rCl}
\capacity(\snr) &\leq &\frac{1}{\cohtime} \sup_{\mimoinpdistclass(\snr)} \mutualinfo{\cinp}{\cout}. \label{ubound_jstar}
\end{IEEEeqnarray}
Note that the right-hand side (RHS) of~\eqref{ubound_jstar} is the capacity of an $\numtxant\times\numrxant$ MIMO Rayleigh bock-fading channel. Hence, it follows from~\cite[Thm. 2]{marzetta1999capacity} that we can restrict the supremum in~\eqref{ubound_jstar} to input distributions for which $\cinp=\bD\bQ,$ where $\bD$ is an $\numtxant\times\numtxant$ real diagonal matrix and $\bQ$ is independent of $\bD,$ and uniformly distributed over the \marktext{Stiefel} manifold $\isotropicgrp(\numtxant,\cohtime).$

Next, we upper-bound the RHS of~\eqref{ubound_jstar} using the duality bound~\cite[Eq. (186)]{lapidoth2003capacity}. This requires the specification of an auxiliary p.d.f. $r_{\cout}(\cdot)$ on $\cout,$ which we choose as follows. Let us denote the singular value decomposition (SVD) of $\cout$ as 
\begin{IEEEeqnarray}{c}
\cout = \bU\bSigma\bV. \label{exp_svd_y}
\end{IEEEeqnarray}
Here, $\bU$ belongs to the set $\unitgrp(\numrxant)$ and has real and positive entries on the main diagonal,\footnote{This second requirement ensures that the SVD in~\eqref{exp_svd_y} is a one-to-one map.} $\bSigma$ is an $ \numrxant\times\numrxant$ diagonal matrix that contains the ordered singular values of $\cout$ on its main diagonal, and $\bV$ belongs to the \marktext{Stiefel} manifold $\isotropicgrp(\numrxant,\cohtime).$ Furthermore, we choose the auxiliary p.d.f. so that the matrices on the RHS of~\eqref{exp_svd_y} are mutually independent, which leads to
\begin{IEEEeqnarray}{rCl}
r_{\cout}(\coutr)&=&f_{\bU}(U) f_{\bSigma}(\Sigma)f_{\bV}(V)J(\Sigma)\label{rydef_split}
\end{IEEEeqnarray} 
where $f_{\bU}(\cdot),\ f_{\bSigma}(\cdot)$ and $f_{\bV}(\cdot)$ are the p.d.f. of $\bU,\ \bSigma$, and $\bV,$ respectively and 
\begin{IEEEeqnarray}{rCl}
J(\Sigma) &=& \frac{1}{\vand(\Sigma^2)^{2}\Bdet{\Sigma}^{2(\cohtime-\numrxant)+1}} \label{ryjacobian}
\end{IEEEeqnarray} is the Jacobian of the SVD~\cite[App. A]{zheng2002communication}.  Moreover, we assume that $\bU$ and $\bV$ are uniformly distributed (with respect to the Haar measure) over their respective domains, which implies that
\begin{IEEEeqnarray}{rCl}
f_{\bU_{\coutr}}(U) &=& \frac{\gammas{\numrxant}}{\pi^{\frac{\numrxant(\numrxant-1)}{2}}} \label{uydist}\\
f_{\bV_{\coutr}}(V) &=& \frac{\gammas{\cohtime}}{\gammas{\cohtime-\numrxant}2^{\numrxant}\pi^{\numrxant\cohtime-\frac{\numrxant(\numrxant-1)}{2}}}. \label{vydist}
\end{IEEEeqnarray}
The p.d.f. of $\bSigma$ is chosen as in~\cite[Sec. V]{6415385}. Specifically, we take the first $\numminant$ singular values of~$\cout$ to be distributed as the singular values of the noiseless channel output $\channel\cinp,$ with $\cinp$ USTM distributed. We take the remaining $\numrxant-\numminant$ singular values of $\cout$ to be independent of the first $\numminant$ singular values and to be distributed as the singular values of an  $(\numrxant-\numminant)\times(\cohtime-\numminant)$  matrix with \iid $\cnormdist{0}{1}$ entries. Let $\bSigma_1$ be an $\numminant\times\numminant$ diagonal matrix containing the first $\numminant$ singular values of $\cout$ and let $\bSigma_2$ be an $(\numrxant-\numminant)\times(\numrxant-\numminant)$ matrix containing the remaining singular values. Then we have \begin{IEEEeqnarray}{rCl}
 f_{\bSigma}(\bSigma_1,\bSigma_2) &=& \frac{ f_{\bSigma_1}(\bSigma_1) f_{\bSigma_2}(\bSigma_2)}{\orderconstant}.
\end{IEEEeqnarray}
Here,
\iftwocol
\begin{IEEEeqnarray}{rCl}
f_{\bSigma_1}(\Sigma_1) &=& \frac{ 2^{\numminant} \gammas{\nummaxant-\numminant}}{\gammas{\numminant}\gammas{\nummaxant}\beta^{\nummaxant\numminant}}\Bexp{-\frac{1}{\beta}\Btrace{\Sigma_1^2}} \nonumber \\
&& \cdot \Bdet{\Sigma_1}^{2(\nummaxant-\numminant)+1}\vand(\Sigma_1^2)^2,\quad \Sigma_1\in\diagmatset{\numminant}{\numminant} \label{sigma1dist}
\end{IEEEeqnarray} 
\else
\begin{IEEEeqnarray}{rCl}
f_{\bSigma_1}(\Sigma_1) &=& \frac{ 2^{\numminant} \gammas{\nummaxant-\numminant}}{\gammas{\numminant}\gammas{\nummaxant}\beta^{\nummaxant\numminant}}\Bexp{-\frac{1}{\beta}\Btrace{\Sigma_1^2}} \Bdet{\Sigma_1}^{2(\nummaxant-\numminant)+1}\vand(\Sigma_1^2)^2,\quad \Sigma_1\in\diagmatset{\numminant}{\numminant} \label{sigma1dist}
\end{IEEEeqnarray} 
\fi
where $\beta = \cohtime\snr/\numtxant$ and 
\iftwocol
\begin{IEEEeqnarray}{rCl}
f_{\bSigma_{2}}(\Sigma_2) &=& \frac{ 2^{\numrxant-\numminant} \gammas{\cohtime-\numrxant}}{\gammas{\numrxant-\numminant}\gammas{\cohtime-\numminant}} e^{-\Btrace{\Sigma_2^2}}\Bdet{\Sigma_2}^{2(\cohtime-\numrxant)+1} \nonumber \\
&&\cdot \vand(\Sigma_2^2)^2,\quad\Sigma_2\in\diagmatset{\numrxant-\numminant}{\numrxant-\numminant}.\IEEEeqnarraynumspace \label{sigma2dist}
\end{IEEEeqnarray}
\else
\begin{IEEEeqnarray}{rCl}
f_{\bSigma_{2}}(\Sigma_2) &=& \frac{ 2^{\numrxant-\numminant} \gammas{\cohtime-\numrxant}}{\gammas{\numrxant-\numminant}\gammas{\cohtime-\numminant}} e^{-\Btrace{\Sigma_2^2}} \Bdet{\Sigma_2}^{2(\cohtime-\numrxant)+1}\vand(\Sigma_2^2)^2,\quad\Sigma_2\in\diagmatset{\numrxant-\numminant}{\numrxant-\numminant}.\IEEEeqnarraynumspace \label{sigma2dist}
\end{IEEEeqnarray}
\fi
Furthermore, $\orderconstant$ is a normalization constant that ensures that the singular values are ordered. Specifically, $\orderconstant$ is the probability that the smallest singular value in $\bSigma_1$ is larger than the largest singular values in $\bSigma_2.$
Using $r_{\cout}(\cdot)$ in the duality bound~\cite[Eq. (186)]{lapidoth2003capacity} we obtain 
\begin{IEEEeqnarray}{rCl}
\capacity(\snr) &\leq &\frac{1}{\cohtime}\ \sup_{\mimoinpdistclass(\snr)}\   \mathopen{}\left\lbrace -{\Bexop{\Blog{r_{\cout}(\cout)}}}-h(\cout\cond\cinp)\right\rbrace \mathclose{}. \label{exp_duality_ry}
\end{IEEEeqnarray}
\marktext{Note that the expectation in~\eqref{exp_duality_ry} is not with respect to~$r_{\cout}(\cdot)$ but with respect to the probability distribution on~$\cout$ induced by the input distribution on $\cinp$ through the channel~\eqref{ysxw}.} Fix $\lambda>0$.
We can further upper-bound the RHS of~\eqref{exp_duality_ry} by using~\eqref{mimo_pwr_const} as follows:
\marktext{\iftwocol
\begin{IEEEeqnarray}{rCl}
\capacity(\snr) &\leq &\frac{1}{\cohtime}\ \inf_{\lambda>0}\sup_{\mimoinpdistclass(\snr)}  \ \mathopen{}\left\lbrace\vphantom{\Bexop{\Btrace{\cinp\cinp^{\dag}}}}-{\Bexop{\Blog{r_{\cout}(\cout)}}}-h(\cout\cond\cinp)\right. \mathclose{}\nonumber\\
&&+\>\mathopen{}\left.\lambda\mathopen{}\left(\cohtime\snr-\Bexop{\Btrace{\cinp\cinp^{\dag}}}\right)\mathclose{}\right\rbrace \mathclose{}. \label{exp_duality_ry_lagrange}
\end{IEEEeqnarray}
\else
\begin{IEEEeqnarray}{rCl}
\capacity(\snr) &\leq &\frac{1}{\cohtime}\ \inf_{\lambda>0}\sup_{\mimoinpdistclass(\snr)}  \ \mathopen{}\left\lbrace-{\Bexop{\Blog{r_{\cout}(\cout)}}}-h(\cout\cond\cinp)+\lambda\mathopen{}\left(\cohtime\snr-\Bexop{\Btrace{\cinp\cinp^{\dag}}}\right)\mathclose{}\right\rbrace \mathclose{}. \label{exp_duality_ry_lagrange}
\end{IEEEeqnarray}
\fi}
Substituting~\eqref{uydist},~\eqref{sigma1dist},~\eqref{sigma2dist},~\eqref{vydist}, and~\eqref{ryjacobian} into~\eqref{exp_duality_ry_lagrange} and using~\cite[Thm. 2]{marzetta1999capacity} we can rewrite the upper bound~\eqref{exp_duality_ry_lagrange} as
\marktext{\iftwocol
\begin{IEEEeqnarray}{rCl}
\capacity(\snr)&\leq & \frac{1}{\cohtime}\ \inf_{\lambda>0}\ \sup_{\mimoinpdistclass(\snr)}   \mathopen{}\left\lbrace\vphantom{\Bexop{\Blog{\frac{\vand(\Sigma^2)^2}{\vand(\Sigma_1^2)^2\vand(\Sigma_2^2)^2}}}} -\numrxant\cohtime+\numrxant\numtxant\Blog{\beta} + \Blog{\orderconstant}\right.\mathclose{}\nonumber \\
&&+\>\Blog{\frac{\gammas{\numrxant-\numminant}\gammas{\cohtime-\numminant}\gammas{\numtxant}}{\gammas{\cohtime} \gammas{\nummaxant-\numminant}}}\nonumber\\
&&-\>\numrxant \Bexop{\Blogdet{I_{\numtxant}+\bD^{2}}}+\lambda\mathopen{}\left(\cohtime\snr-\Bexop{\Btrace{\bD^{2}}}\right)\mathclose{}\nonumber\\
&&+\>\frac{1}{\beta}\Bexop{\Btrace{\bSigma_1^2}}+\Bexop{\Btrace{\bSigma_2^2}}\nonumber \\
&&+\>(\numminant+\cohtime-\nummaxant-\numrxant)\Bexop{\Blogdet{\bSigma_1^2}} \nonumber\\
&&+\>\mathopen{}\left.\Bexop{\Blog{\frac{\vand(\Sigma^2)^2}{\vand(\Sigma_1^2)^2\vand(\Sigma_2^2)^2}}} \right\rbrace\mathclose{}.\IEEEeqnarraynumspace \label{ub_long_exp}
\end{IEEEeqnarray} 
\else
\begin{IEEEeqnarray}{rCcl}
\capacity(\snr)&\leq & \frac{1}{\cohtime}\ \inf_{\lambda>0}\ \sup_{\mimoinpdistclass(\snr)}  & \left\lbrace -\numrxant\cohtime+\numrxant\numtxant\Blog{\beta} + \Blog{\orderconstant}+\Blog{\frac{\gammas{\numrxant-\numminant}\gammas{\cohtime-\numminant}\gammas{\numtxant}}{\gammas{\cohtime} \gammas{\nummaxant-\numminant}}}\right.\nonumber\\
&&&-\>\numrxant \Bexop{\Blogdet{I_{\numtxant}+\bD^{2}}}+\lambda\mathopen{}\left(\cohtime\snr-\Bexop{\Btrace{\bD^{2}}}\right)\mathclose{}\nonumber\\
&&&+\>\frac{1}{\beta}\Bexop{\Btrace{\bSigma_1^2}}+\Bexop{\Btrace{\bSigma_2^2}}+(\numminant+\cohtime-\nummaxant-\numrxant)\Bexop{\Blogdet{\bSigma_1^2}} \nonumber\\
&&&+\>\mathopen{}\left.\Bexop{\Blog{\frac{\vand(\Sigma^2)^2}{\vand(\Sigma_1^2)^2\vand(\Sigma_2^2)^2}}} \right\rbrace\mathclose{}.\IEEEeqnarraynumspace \label{ub_long_exp}
\end{IEEEeqnarray} 
\fi }
We proceed now by upper-bounding some of the terms on the RHS of~\eqref{ub_long_exp}.
First we bound $\Bexop{\Btrace{\bSigma_1^2}}$:
\begin{IEEEeqnarray}{rCcCl}
\Bexop{\Btrace{\bSigma_1^2}}&\leq & \Bexop{\Btrace{\cout\cout^{\dag}}}
&=& \numrxant\cohtime+\numrxant\Bexop{\Btrace{\bD^{2}}}. \IEEEeqnarraynumspace \label{boundsigma1sq}
\end{IEEEeqnarray}
Let $\sigma_1^2>\sigma_2^2>\dots>\sigma_{\numtxant}^2$ denote the eigenvalues of $\cout\cout^{\dag}.$ To bound $\Bexop{\Btrace{\bSigma_2^2}},$ we use the argument given in~\cite[p. 377]{zheng2002communication}, which yields 
\begin{IEEEeqnarray}{rCcCl}
\Bexop{\Btrace{\bSigma_2^2}} &=&\Bexop{\sum_{i=\numminant+1}^{\numrxant}{\sigma_i^2}}&\leq & (\numrxant-\numminant)(\cohtime-\numminant). \label{boundsigma2sq}
\end{IEEEeqnarray}
We also have that 
\iftwocol
\begin{IEEEeqnarray}{rCl}
\Bexop{\Blog{\frac{\vand(\Sigma^2)^2}{\vand(\Sigma_1^2)^2\vand(\Sigma_2^2)^2}}} &=& \Bexop{\sum_{i=1}^{\numminant}\sum_{j=\numminant+1}^{\numrxant}2\Blog{\sigma_i^2-\sigma_j^2}} \nonumber\\
 &\leq & 2(\numrxant-\numminant)\Bexop{ \Blogdet{\bSigma_1^2}}. \IEEEeqnarraynumspace
\end{IEEEeqnarray}
\else
\begin{IEEEeqnarray}{rCcCcl}
\Bexop{\Blog{\frac{\vand(\Sigma^2)^2}{\vand(\Sigma_1^2)^2\vand(\Sigma_2^2)^2}}} &=& \Bexop{\sum_{i=1}^{\numminant}\sum_{j=\numminant+1}^{\numrxant}2\Blog{\sigma_i^2-\sigma_j^2}} 
 &\leq & 2(\numrxant-\numminant)\Bexop{ \Blogdet{\bSigma_1^2}}. \IEEEeqnarraynumspace
\end{IEEEeqnarray}
\fi
Finally, to bound the term $\Bexop{ \Blogdet{\bSigma_1^2}},$ we start by noting that, given $\cinp=X=DQ,$ the rows of the random matrix $\cout$ are \iid complex Gaussian random vectors with zero mean and covariance matrix $I_{\cohtime}+X^\dag X.$ This means that, given $\cinp=X=DQ,$ the matrix $\cout^\dag\cout$ has the same distribution as  $\bG(I_{\numtxant}+D^2)\bG^{\dag} + \bH\bH^{\dag}$ where $\bG\in \complexset^{\numrxant\times \numtxant}$ and $\bH\in \complexset^{\numrxant\times(\cohtime-\numtxant)}$ are independent matrices with \iid $\cnormdist{0}{1}$ entries. Let $\lambda_1,\lambda_2,\dots,\lambda_{\numminant}$ denote the $\numminant$ eigenvalues of $\bG(I_{\numtxant}+D^2)\bG^{\dag}$ (note that this matrix has rank~$\numminant$). Furthermore, let $\mu_1$ denote the largest eigenvalue of $\bH\bH^{\dag},$ and let $\expmaxaigenconstant = \Bexop{\mu_1}.$ We have 
\marktext{\iftwocol
\begin{IEEEeqnarray}{rCl}
\Bexop{\Blogdet{\bSigma_1^2}} &=& \Bexop{\sum_{i=1}^{\numminant}\Blog{\sigma_i^2}}\\
&=& \Bexoprand{\bD,\bQ}{\Bexoprand{\cout\cond\bD,\bQ}{\sum_{i=1}^{\numminant}\Blog{\sigma_i^2}\middle\cond \bD,\bQ }}\IEEEeqnarraynumspace\\
&\leq & \Bexoprand{\bD}{\Bexoprand{\bG, \bH}{\sum_{i=1}^{\numminant}\Blog{\lambda_i+\mu_1}\middle\cond \bD}} \label{weylstep}\\
&\leq & \Bexoprand{\bD}{\Bexoprand{\bG}{\sum_{i=1}^{\numminant}\Blog{\lambda_i+\expmaxaigenconstant}\middle\cond \bD}} \label{jensenstep}\\
&=& \Bexop{\Blogdet{\bG(I_{\numtxant}+\bD^2)\bG^{\dag}+\expmaxaigenconstant I_{\numrxant}}} \nonumber \\
&&-\> (\numrxant-\numminant)\Blog{\expmaxaigenconstant}.\label{rankstep}
\end{IEEEeqnarray}
\else
\begin{IEEEeqnarray}{rCl}
\Bexop{\Blogdet{\bSigma_1^2}} &=& \Bexop{\sum_{i=1}^{\numminant}\Blog{\sigma_i^2}}\\
&=& \Bexoprand{\bD,\bQ}{\Bexoprand{\cout\cond\bD,\bQ}{\sum_{i=1}^{\numminant}\Blog{\sigma_i^2}\middle\cond \bD,\bQ }}\\
&\leq & \Bexoprand{\bD}{\Bexoprand{\bG, \bH}{\sum_{i=1}^{\numminant}\Blog{\lambda_i+\mu_1}\middle\cond \bD}} \label{weylstep}\\
&\leq & \Bexoprand{\bD}{\Bexoprand{\bG}{\sum_{i=1}^{\numminant}\Blog{\lambda_i+\expmaxaigenconstant}\middle\cond \bD}} \label{jensenstep}\\
&=& \Bexop{\Blogdet{\bG(I_{\numtxant}+\bD^2)\bG^{\dag}+\expmaxaigenconstant I_{\numrxant}}} - (\numrxant-\numminant)\Blog{\expmaxaigenconstant}.\label{rankstep}
\end{IEEEeqnarray}
\fi }
Here,~\eqref{weylstep} follows from Weyl's theorem~\cite[Thm. 4.3.1]{horn1990matrix} and in~\eqref{jensenstep} we used Jensen's inequality.
Substituting~\eqref{boundsigma1sq},~\eqref{boundsigma2sq},~\eqref{rankstep} into~\eqref{ub_long_exp} we obtain
 \begin{IEEEeqnarray}{rCl}
\capacity(\snr)&\leq& \marktext{u(\snr)} + \frac{1}{\cohtime}\inf_{\lambda\geq 0}\sup_{ \mimoinpdistclass(\snr)} \Bexop{g(\bD,\lambda)} \label{ub_exp_before_sup}\\
&\leq& \marktext{u(\snr)} + \frac{1}{\cohtime}\inf_{\lambda\geq 0}  \sup_{D\in \diagmatset{\numtxant}{\numtxant}} g(D,\lambda)\label{ub_exp_sup}
\end{IEEEeqnarray} where $u(\cdot)$ and $g(\cdot,\cdot)$   are given in~\eqref{ub_udef} and~\eqref{ub_gdef}, respectively, and where the last step follows by upper-bounding the supremum over the probability distribution on $\bD$ with the supremum over the set of deterministic diagonal matrices D. 
% We conclude the proof by noting that
% \begin{IEEEeqnarray}{c}
% \Bexop{g(\bD,\lambda)} \leq \sup_{D\in \diagmatset{\numtxant}{\numtxant}} g(D,\lambda)
% \end{IEEEeqnarray} which substituted into~\eqref{ub_exp_before_sup} yields~\eqref{ub_expr}.

\section{Proof of Corollary~\ref{theorem_ub_sym}}
\label{appendix_ub_symproof}
When $\numtxant=\numrxant=\numsymant,$ the upper bound~\eqref{ub_long_exp} becomes
\marktext{\iftwocol
\begin{IEEEeqnarray}{rCl}
\capacity(\snr)&\leq & \frac{1}{\cohtime}\ \inf_{\lambda>0}\ \sup_{\mimoinpdistclass(\snr)}   \left\lbrace \numsymant^{2}\Blog{\beta} +\Blog{\frac{\gammas{\numsymant}\gammas{\cohtime-\numsymant}}{\gammas{\cohtime}}}\right.\nonumber \\
&&-\>\numsymant\cohtime+(\cohtime-\numsymant)\Bexop{\Blogdet{\cout\cout^{\dag}}}\nonumber \\
&&-\>\numsymant\Bexop{\Blogdet{I_\numsymant+\bD^{2}}}\nonumber\\
&&+\>\mathopen{}\left.\frac{1}{\beta}\Bexop{\Btrace{\cout\cout^{\dag}}}+\lambda\mathopen{}\left(\cohtime\snr-\Bexop{\Btrace{\bD^{2}}}\right)\mathclose{}\vphantom{\Blog{\frac{\gammas{\numsymant}\gammas{\cohtime-\numsymant}}{\gammas{\cohtime}}}}\right\rbrace\mathclose{}. \IEEEeqnarraynumspace
\label{ub_long_exp_sym}
\end{IEEEeqnarray}
\else
\begin{IEEEeqnarray}{rCcl}
\capacity(\snr)&\leq & \frac{1}{\cohtime}\ \inf_{\lambda>0}\ \sup_{\mimoinpdistclass(\snr)}  & \left\lbrace -\numsymant\cohtime+\numsymant^{2}\Blog{\beta} +\Blog{\frac{\gammas{\numsymant}\gammas{\cohtime-\numsymant}}{\gammas{\cohtime}}}\right.\nonumber \\
&&&+\>(\cohtime-\numsymant)\Bexop{\Blogdet{\cout\cout^{\dag}}}-\numsymant\Bexop{\Blogdet{I_\numsymant+\bD^{2}}}\nonumber\\
&&&+\>\mathopen{}\left.\frac{1}{\beta}\Bexop{\Btrace{\cout\cout^{\dag}}}+\lambda\mathopen{}\left(\cohtime\snr-\Bexop{\Btrace{\bD^{2}}}\right)\mathclose{}\right\rbrace\mathclose{}. \IEEEeqnarraynumspace
\label{ub_long_exp_sym}
\end{IEEEeqnarray}
\fi }
Differently from the general case treated in Appendix~\ref{appendix_ubproof}, in the square case some of the terms on the RHS of~\eqref{ub_long_exp_sym} can be computed in closed form. Specifically,
\begin{IEEEeqnarray}{rCl}
\Bexop{\Btrace{\cout\cout^{\dag}}} &=& \numsymant\cohtime+\numsymant\Bexop{\Btrace{\bD^2 }}. \label{expyyh_sym}
\end{IEEEeqnarray}
To evaluate $\Bexop{\Blogdet{\cout\cout^{\dag}}}$ we use that, given $\cinp=X=DQ,$ the rows of $\cout$ are \iid complex Gaussian random vectors with zero mean and covariance matrix $I_{\cohtime}+X^{\dag}X.$ This means that given $\cinp=X=DQ,$ the matrix $\cout\cout^{\dag}$ has the same probability distribution as $\bZ(I_{\cohtime}+\cinpr^\dag \cinpr)\bZ^{\dag},$ where~$\bZ\in \complexset^{\numsymant\times \cohtime}$ has \iid $\cnormdist{0}{1}$ entries. Hence, we conclude that 
\marktext{\iftwocol
\begin{IEEEeqnarray}{rCl}
\IEEEeqnarraymulticol{3}{l}{\Bexop{\Blogdet{\cout\cout^{\dag}}} } \nonumber \\
&=&\Bexoprand{\cinp}{\Bexoprand{\bZ}{\Blogdet{\bZ(I_{\cohtime}+\cinp^\dag \cinp)\bZ^{\dag}}}}\\
&=& \Bexop{\frac{ \Bdet{I_\numsymant+\bD^{2}}^{\cohtime-\numsymant-1}}{\kappas{\bD^{2}}{\cohtime-\numsymant}}  \sum_{k=1}^{\numsymant}\Bdet{R_k(I_\numsymant+\bD^{2})}}.\IEEEeqnarraynumspace \label{explogdet_sym}
\end{IEEEeqnarray}%
\else
\begin{IEEEeqnarray}{rCl}
\Bexop{\Blogdet{\cout\cout^{\dag}}}  &=& \Bexoprand{\cinp}{\Bexoprand{\bZ}{\Blogdet{\bZ(I_{\cohtime}+\cinp^\dag \cinp)\bZ^{\dag}}}}\\
&=& \Bexop{\frac{ \Bdet{I_\numsymant+\bD^{2}}^{\cohtime-\numsymant-1}}{\kappas{\bD^{2}}{\cohtime-\numsymant}}  \sum_{k=1}^{\numsymant}\Bdet{R_k(I_\numsymant+\bD^{2})}}. \label{explogdet_sym}
\end{IEEEeqnarray}
\fi }
Here, the last step follows from Lemma~\ref{lem_exp_log_det_gauss} in Appendix~\ref{appendix_pre_result}. Substituting~\eqref{expyyh_sym} and~\eqref{explogdet_sym} into~\eqref{ub_long_exp_sym} we obtain 
 \begin{IEEEeqnarray}{c}
\capacity(\snr)\leq \marktext{u^{*}(\snr)} + \frac{1}{\cohtime}\inf_{\lambda\geq 0}\sup_{ \mimoinpdistclass(\snr)} \Bexop{g^{*}(\bD,\lambda)}\label{ub_exp_sym_sup}
\end{IEEEeqnarray} where $ u^{*}(\cdot)$ and $g^{*}(\cdot,\cdot)$ are given in~\eqref{ub_udef_sym} and~\eqref{ub_gdef_sym} respectively. We conclude the proof by noting that
\begin{IEEEeqnarray}{c}
\Bexop{g^{*}(\bD,\lambda)} \leq \sup_{D\in \diagmatset{\numsymant}{\numsymant}} g^{*}(D,\lambda)
\end{IEEEeqnarray}
which yields~\eqref{expr_ub_sym}.

\section{ Proof of Theorem~\ref{theorem_lb}}
\label{appendix_lb_proof}
We lower-bound $\capacity(\snr)$ by evaluating the mutual information on the RHS of~\eqref{sumratecap} for the MAC-USTM distribution~\eqref{spunidist}, which yields \begin{IEEEeqnarray}{c}
\capacity(\snr) \geq \mutualinfo{\cinp}{\cout}.
\end{IEEEeqnarray} Let $\cinp = \bU \bD \bQ$ denote the singular value decomposition of $\cinp,$ where $\bU$ belongs to the unitary group $\unitgrp(\numtxant),$ the diagonal matrix $\bD\in\diagmatset{\numtxant}{\numtxant}$ contains the ordered singular values of $\cinp$ on its main diagonal, and $\bQ$ belongs to the \marktext{Stiefel} manifold $\isotropicgrp(\numtxant,\cohtime)$.  We next decompose the mutual information as \begin{IEEEeqnarray}{rCl}
\mutualinfo{\cinp}{\cout} &=&\difent(\cout)-\difent(\cout\cond\cinp) \label{ixy_hy_hy_x}
\end{IEEEeqnarray} and evaluate the two terms on the RHS of~\eqref{ixy_hy_hy_x}  separately. Since $\cout$ is conditionally Gaussian given $\cinp,$ the second term in~\eqref{ixy_hy_hy_x} can be simplified as \begin{IEEEeqnarray}{rCl}
\difent(\cout\cond\cinp) &=&\numrxant\Bexop{\Blogdet{I_{\numtxant}+\bD^2}} + \numrxant\cohtime\Blog{\pi e}.\IEEEeqnarraynumspace\label{exprhygivenx}
\end{IEEEeqnarray} 
We next evaluate $\difent(\cout).$ Since $\cinp V\sim\cinp$ for every deterministic unitary matrix $V\in\unitgrp(\cohtime),$ we conclude that $\bQ$ is uniformly distributed over $\isotropicgrp(\numtxant,\cohtime)$ and independent of $\bD.$ We also have that \begin{IEEEeqnarray}{rCl}
\cout &=& \channel\cinp+\noise\\
&=&\channel\bU \bD \bQ+\noise \label{y_sudq_w_step}\\
&\sim&\channel\bD \bQ+\noise.\label{y_udq_w_step}
\end{IEEEeqnarray} Here,~\eqref{y_sudq_w_step} follows from the singular value decomposition of $\cinp,$ and~\eqref{y_udq_w_step} holds because $\channel$ is isotropically distributed, and, hence, $\channel\bU \sim \channel$, which also implies that $\channel\bU$ is independent of $\bD$. The fact that $\bQ$ is uniformly distributed over~$\isotropicgrp(\numtxant,\cohtime)$ and independent of $\bD$ allows us to invoke \cite[Prop. 6]{alfano2014closed} and obtain a closed form expression for the conditional p.d.f. $f_{\cout\cond\bD}$ of $\cout$ given $\bD.$ Substituting this closed form expression into \begin{IEEEeqnarray}{rCl}
\difent(\cout)&=&-{\Bexop{\Blog{f_{\cout}(\cout)}}}\\
 &=& -{\Bexop{\Blog{\Bexoprand{\bD}{f_{\cout\cond\bD}(\cout\cond\bD)}}}}\label{exprhy_ln_exp_y_d}
\end{IEEEeqnarray} and then substituting~\eqref{exprhy_ln_exp_y_d} and~\eqref{exprhygivenx} into~\eqref{ixy_hy_hy_x}, we obtain~\eqref{exprixy}.

\section{Proof of Corollary~\ref{cor_lb_2x2}}
\label{appendix_lb_2x2_proof}
We derive in closed form the p.d.f. of the eigenvalues of the $2\times 2$ matrix $\cinp\cinp^{\dag},$ where the two rows $\cinp_1\in\complexset^{1\times\cohtime}$ and $\cinp_2\in\complexset^{1\times\cohtime}$ of the $2\times\cohtime$ dimensional matrix $\cinp$ are given by \begin{IEEEeqnarray}{rCl}
\cinp_i &=& \sqrt{\frac{\cohtime \snr}{2}}\bV_i, \quad i=1,2
\end{IEEEeqnarray} with $\bV_1\in\complexset^{1\times\cohtime}$ and $\bV_2\in\complexset^{1\times\cohtime}$ \iid and uniformly distributed on $\isotropicgrp(1,\cohtime)$ (with respect to the Haar measure). 
To compute the eigenvalues of $\cinp\cinp^{\dag},$ we express it as follows  \begin{IEEEeqnarray}{rCl}
\cinp\cinp^{\dag} &=& \frac{\cohtime\snr}{2} \begin{bmatrix}
\bV_1 \\
\bV_2
\end{bmatrix} \begin{bmatrix}
\bV_1^{\dag} & \bV_2^{\dag}
\end{bmatrix}\\
&=& \frac{\cohtime\snr}{2} \begin{bmatrix}
1& \bV_1\bV_2^{\dag} \\
\bV_2\bV_1^{\dag} & 1
\end{bmatrix}\marktext{.} \label{expr2x2_xxh}
\end{IEEEeqnarray} Let $\alpha$ denote the absolute value of the scalar $\bV_1\bV_2^{\dag}.$ It follows from~\eqref{expr2x2_xxh} that the eigenvalues of $\cinp\cinp^{\dag}$ are $\cohtime\snr(1\pm\alpha)/2.$ Hence, in order to determine the p.d.f. of the eigenvalues of $\cinp\cinp^{\dag},$ it is sufficient to obtain the p.d.f. of $\alpha.$ Since \begin{IEEEeqnarray}{rCl}
\Bdet{\cinp\cinp^{\dag}} &=& \left( \frac{\cohtime\snr}{2}\right)^{2}(1-\alpha^{2})
\end{IEEEeqnarray} Bartlett's decomposition (see~\cite[Prop. 2.1]{rouault2006asymptotic}) implies that  $\alpha^2 \sim  \betadist{1}{\cohtime-1}.$ After some mathematical manipulations, we obtain that \begin{IEEEeqnarray}{rCl}
f_{\alpha}(x) &=&(\cohtime-1)2x(1-x^2)^{\cohtime-2},\quad 0\leq x \leq 1. \label{alppdf}
\end{IEEEeqnarray} Using~\eqref{alppdf} in~\eqref{exprixy} together with the expression for the eigenvalues of $\cinp\cinp^{\dag},$ we obtain~\eqref{exprixy2case}.
\end{appendices}

\bibliographystyle{IEEEtran}
\bibliography{IEEEabrv,confs-jrnls,publishers,refs}

\end{document}